\documentclass[AMA,Times1COL]{WileyNJDv5} 

\articletype{Article Type}%

\received{Date Month Year}
\revised{Date Month Year}
\accepted{Date Month Year}
\journal{Journal}
\volume{00}
\copyyear{2023}
\startpage{1}

\usepackage{booktabs}
\usepackage{subfig}
\usepackage{longtable}
\usepackage{hyperref}

\begin{document}

\title{Interconnections of Multimorbidity-Related Clinical Outcomes: Analysis of Health Administrative Claims Data with a Dynamic Network Approach}

\author[1,2,3]{Hao Mei}
\author[2]{Haonan Xiao}
\author[4]{Ben-Chang Shia}
\author[5]{Guanzhong Qiao}
\author[1,2,3]{Yang Li}

\authormark{TAYLOR \textsc{et al.}}
\titlemark{Interconnections of Multimorbidity-Related Clinical Outcomes: Analysis of Health Administrative Claims Data with a Dynamic Network Approach}

\address[1]{\orgdiv{Center for Applied Statistics}, \orgname{Renmin University of China}, \orgaddress{\state{Beijing}, \country{China}}}
\address[2]{\orgdiv{School of Statistics}, \orgname{Renmin University of China}, \orgaddress{\state{Beijing}, \country{China}}}
\address[3]{\orgdiv{Institute of Health Data Science}, \orgname{Renmin University of China}, \orgaddress{\state{Beijing}, \country{China}}}
\address[4]{\orgdiv{Graduate Institute of Business Administration, College of Management; Artificial Intelligence Development Center}, \orgname{Fu Jen
Catholic University}, \orgaddress{\state{Taipei}, \country{Taiwan}}}
\address[5]{\orgdiv{Department of Orthopaedic}, \orgname{The First Hospital of Tsinghua University}, \orgaddress{\state{Beijing}, \country{China}}}

\corres{Corresponding author Yang Li, School of Statistics, Renmin University of China, Beijing 100872, China \email{yang.li@ruc.edu.cn}}



\abstract{Given the rising complexity and burden of multimorbidity, it is crucial to provide evidence-based support for managing multimorbidity-related clinical outcomes. This study introduces a dynamic network approach to investigate conditional and time-varying interconnections in disease-specific clinical outcomes. Our method effectively tackles the issue of zero inflation, a frequent challenge in medical data that complicates traditional modeling techniques. The theoretical foundations of the proposed approach are rigorously developed and validated through extensive simulations. Using Taiwan's health administrative claims data from 2000 to 2013, we construct 14 yearly networks that are temporally correlated, featuring 125 nodes that represent different disease conditions. Key network properties, such as connectivity, module, and temporal variation are analyzed. To demonstrate how these networks can inform multimorbidity management, we focus on breast cancer and analyze the relevant network structures. The findings provide valuable clinical insights that enhance the current understanding of multimorbidity. The proposed methods offer promising applications in shaping treatment strategies, optimizing health resource allocation, and informing health policy development in the context of multimorbidity management.}


\jnlcitation{\cname{%
\author{Taylor M.},
\author{Lauritzen P},
\author{Erath C}, and
\author{Mittal R}}.
\ctitle{On simplifying ‘incremental remap’-based transport schemes.} \cjournal{\it J Comput Phys.} \cvol{2021;00(00):1--18}.}

\maketitle



\section{Introduction}
Multimorbidity, defined as the presence of two or more coexisting medical conditions in an individual, affects an estimated 37.2\% of adults worldwide, a prevalence that is expected to continue increasing due to aging populations and lifestyle factors. \citep{chowdhury2023global} Medical costs associated with multimorbidity have been reported to range from \$800 to \$150,000 per person, depending on the specific disease combinations and other factors. \citep{tran2022costs} Multimorbidity is emerging as a critical health issue, imposing significant burdens on individual patients, the healthcare system, and society.
However, existing studies on multimorbidity primarily focus on the mechanisms of disease development \citep{Tazzeo2023,zhang2024disease} or on integrated care models targeting coexisting diseases. \citep{rohwer2023models, tops2024integrated} Yet, clinical outcomes, which are direct measurements of healthcare effectiveness and efficiency, have not been sufficiently addressed in multimorbidity research.\\


Clinical outcomes include healthcare results that patients can directly perceive (e.g., cure, death, and quality of life) and the measures of resources spent to achieve theses results (e.g., number of outpatient visits, inpatient length of stay, and medical costs). Prior research on clinical outcomes generally focuses on a single disease, and therefore cannot adequately inform multimorbidity management. There are a few multimorbidity-related clinical outcome studies, but are limited to a small number of pre-selected diseases or all diseases combined results. For example, Mariotto et al. (2020) \citep{mariotto2020medical} estimated cancer-attributable medical costs based on the Surveillance, Epidemiology, and End Results database, controlling for the presence of multiple pre-specified chronic diseases; Zeng et al. (2021) \citep{zeng2021multi} offered insights into deep learning methodologies for predicting individual-level all diseases combined medical costs with historical claims data collected from the Medicaid program. Such studies have a limited scope and cannot provide the ``big picture'' of the complex disease interconnections affecting clinical outcomes.\\


Human Disease Network (HDN) analysis is a powerful tool for studying disease interconnections from a ``global'' perspective. Earlier HDNs primarily focus on shared molecular risk factors \citep{goh2007human, silverman2020molecular} or phenotypic correlations between diseases, \citep{hidalgo2009dynamic, jiang2018epidemiological} without incorporating clinical outcome information.
Recent developments in HDNs have expanded the analysis to investigate intercorrelations of disease-specific clinical outcomes. Compared to molecular and phenotypic HDNs, clinical outcome HDNs are directly relevant to clinical practice and may have greater practical value. For example, Ma et al. (2020) \citep{ma2020human}constructed HDNs analyzing unconditional pairwise correlations in disease-specific medical costs. Mei et al. (2021, 2023) \citep{mei2021human, mei2023human} further advanced HDN analysis by focusing on conditional disease interconnections across multiple outcomes. Unconditional HDNs determine pairwise disease connections ignoring the presence of other diseases, whereas conditional HDNs assess these connections accounting for the influence of other diseases. The two types of analyses may differ in scope and interpretation, with conditional analysis being methodologically more complex.\\

Although existing multimorbidity research, particularly advances in HDN, have significantly improved our understanding of how coexisting diseases interact and influence clinical outcomes, these methods face methodological limitations when applied to health administrative claims data. Health administrative claims data, also referred as claims data, are records generated through the administration of health insurance claims. The rapid accumulation of such data presents an opportunity for more comprehensive multimorbidity research on clinical outcomes. 
This type of data has unique advantages. Firstly, it provides extensive coverage of the insured population. However, it is important to recognize the potential for selection bias due to variations in actual insurance coverage. As a result, careful consideration of research questions is crucial to minimize this limitation. Second, claims data often include comprehensive information about the health of insurance beneficiaries, such as diagnosis, admissions, prescriptions, operations, mortality, and costs. Third, since claim data originate from routine healthcare services, studies leveraging such data are generally more efficient in evaluating the real-world effectiveness, safety, and costs of health interventions. Finally, since claim data are used for insurance reimbursement, the data quality is generally high. Claims data have been extensively studied, as seen in the aforementioned mentioned multimorbidity studies \citep{tran2022costs, Tazzeo2023} and clinical outcome studies. \citep{mariotto2020medical, rohwer2023models} However, conducting multimorbidity-related clinical outcome studies based on claims data poses several challenges. Specifically, modeling the interconnections between disease-specific clinical outcomes while accounting for zero inflation and temporal variation requires additional methodological developments.\\

\begin{figure}[htbp]
    \centering
    \captionsetup{justification=centering}
    \subfloat[Joint and marginal distributions of medical costs of Essential Hypertension and Chronic Obstructive Pulmonary Disease]{
    	\includegraphics[width=0.6\textwidth]{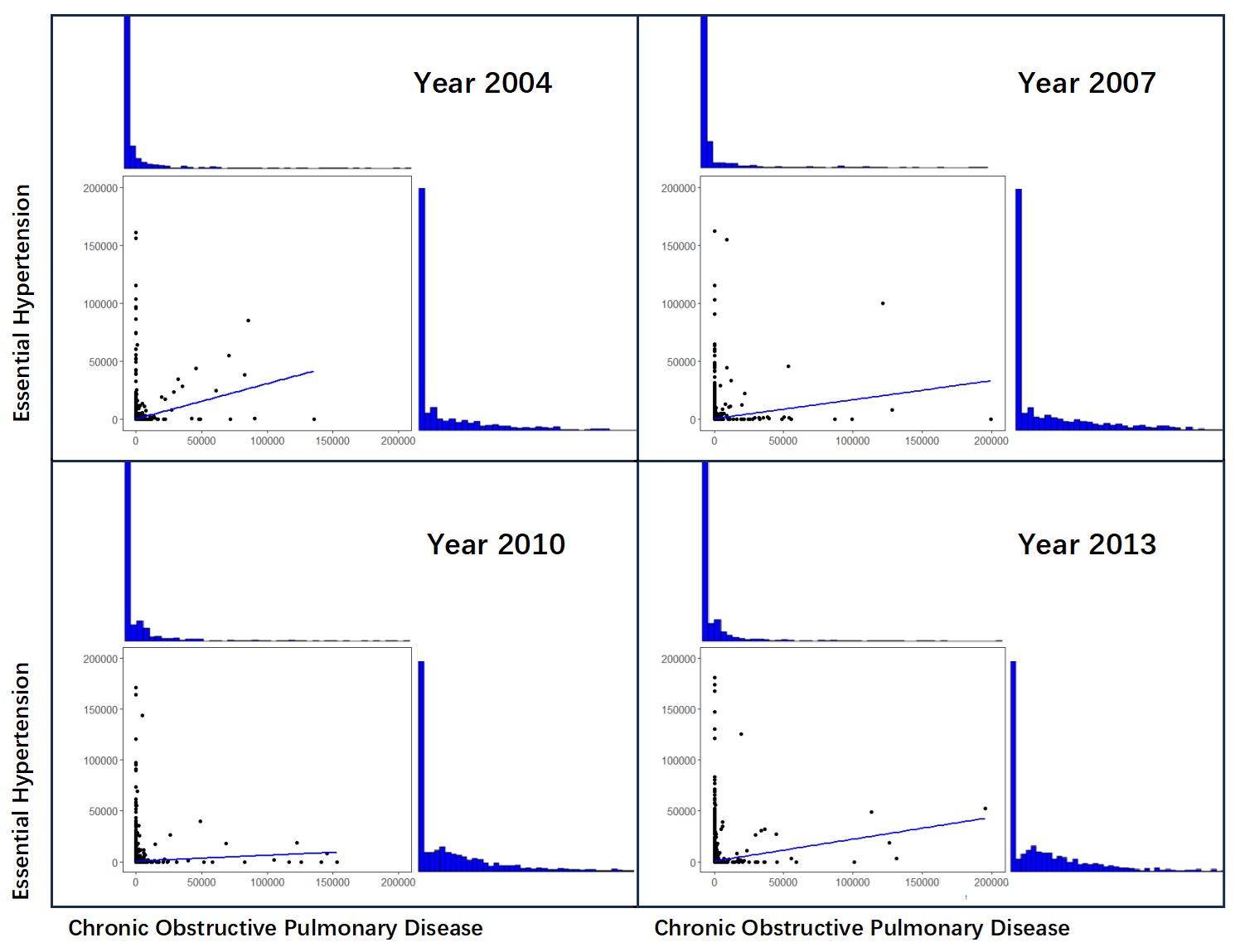} 
     \label{fig:joint_marginal}
     }
     
    \subfloat[Heatmap of pairwise Pearson correlations of disease-specific medical costs]{
    	\includegraphics[width=0.465\textwidth]{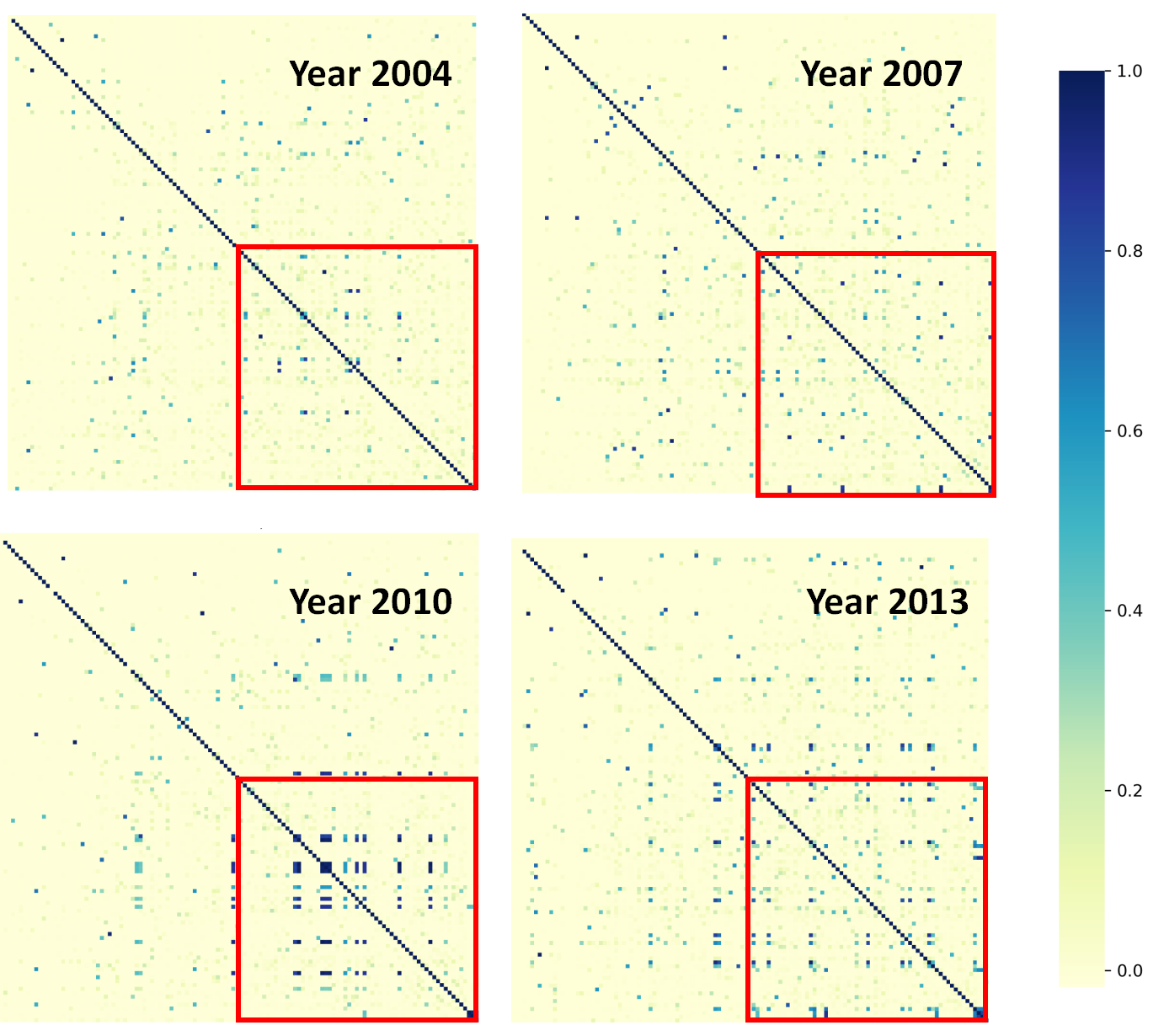} 
     \label{fig:heatmap}
     }
     \subfloat[Enlarged portions of Figure 1(b)]{
    	\includegraphics[width=0.4\textwidth]{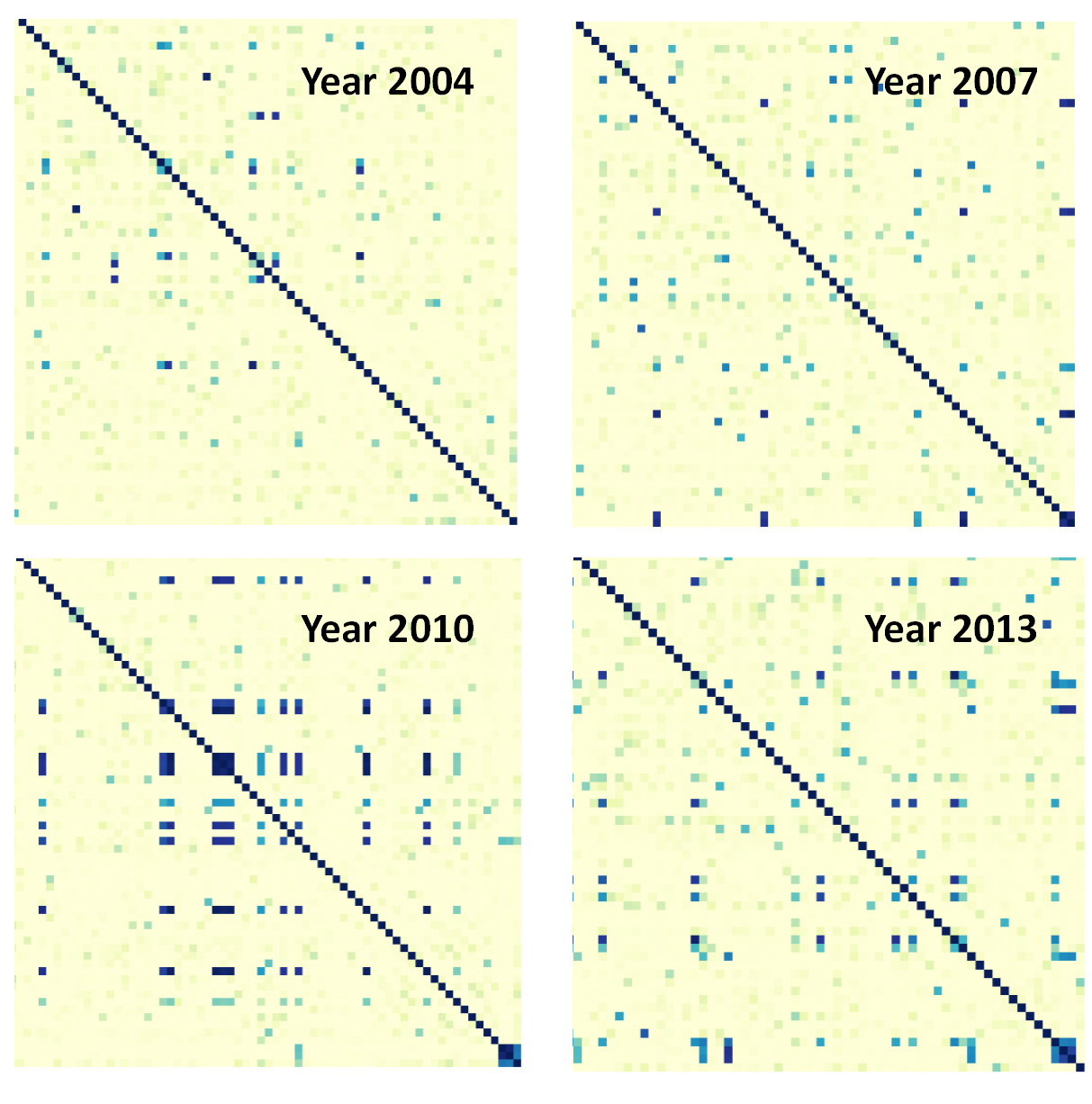}
     \label{fig:heatmap_enlarge}
     }
     
    \caption{Descriptive figures of disease-specific medical costs over different years}
    \label{fig:discriptive}
\end{figure}

In this study, we analyze medical costs data from Taiwan's claims records for the period 2000 to 2013. 
Figure \ref{fig:discriptive} presents descriptive results of disease-specific medical costs for 125 diseases over four different years (see Section 4.1 for detailed data processing). Figure \ref{fig:joint_marginal} displays the joint and marginal distributions of medical costs for two common diseases, Essential Hypertension and Chronic Obstructive Pulmonary Disease. The nonparametric fit (blue lines in the scatter plots) demonstrates a positive correlation in all four years, with the correlation's magnitude varying across years. Moreover, the marginal distributions clearly reveal the zero-inflation characteristic of the data. 
In our analysis, we focus on total medical costs, a validated entry in Taiwan's claims data that encompasses both deductible and reimbursed amounts. Zero costs in the dataset can arise from two distinct scenarios. The first pertains to healthy beneficiaries who do not seek treatment for specific diseases. Existing studies have highlighted that clinical outcome data at the pan-disease level are generally zero-inflated, as the population is a mix of both healthy individuals and those affected by specific diseases. \citep{mei2023human,liu2019statistical} The second scenario involves diseased individuals who have not sought treatment through the insurance program for various reasons. This limitation underscores a key constraint of claims data, as it only captures healthcare utilization covered by the insurance scheme. Given the zero-inflation characteristic of the data, it is important to note that most existing multimorbidity studies focus on binary and continuous outcomes, which may not sufficiently address the complexities of zero-inflated data.
\citep{mariotto2020medical,zeng2021multi}
Figure \ref{fig:heatmap} shows the heatmap of pairwise Pearson correlations of medical costs for all 125 diseases. A significant number of disease pairs exhibit moderate to strong correlations, underscoring the importance of studying disease interconnections in clinical outcomes at the pan-disease level. Figure \ref{fig:heatmap_enlarge}, enlarged portions of Figure \ref{fig:heatmap}, highlights that the correlation structure varies over time. Various factors, such as changes in social environments, advancements in medical technology, and developments in health policies, can lead to the temporal evolution of disease interconnections, altering the risk structure of clinical outcomes. \citep{wallace2022epidemiology} This dynamic feature further complicates the analysis. 
While some HDN analysis \citep{hidalgo2009dynamic, ma2020human, mei2023human} investigate temporal trends by constructing and comparing static networks year by year, this approach lacks methodological and theoretical justification, compromising validity of the inferences drawn.\\

To address challenges in studying multimorbidity-related clinical outcomes with claims data, we propose a dynamic HDN for studying disease interconnections in medical costs. In this network, one node corresponds to a disease, and two nodes are linked with an edge if their medical costs are conditionally correlated under the proposed model. We select medical costs, a direct measure of disease severity and disease attributable financial burden, as the targeted clinical outcome because of its high clinical relevance. It is noted that the proposed model can be easily extended to other outcomes or cases of multiple outcomes. To address zero-inflation, we adopt a two-part modeling approach that separately estimates the probability and magnitude of non-zero medical costs. To capture temporal evolution, we consider time-dependent and smoothed coefficients. To address theoretical gaps in HDN studies, we establish the identifiability of the model and demonstrate the consistency of estimation and variable selection. The proposed model is applied to Taiwan's claims data, expanding the scope of analysis and application of administrative claims data. Based on the constructed network, we analyze key network properties, including connectivity, module, and temporal variation. Findings can enhance decision-making in multimorbidity management. As suggested in published HDN studies, \citep{hidalgo2009dynamic,ma2020human,mei2023human} interventions targeting diseases connected to a large number of others or groups of tightly interconnected diseases could be more efficient and cost-effective.\\


\section{Methods}
To analyze conditional and time-varying disease interconnections in medical costs, we 
propose a dynamic graphical model incorporating two-part modeling and integrative analysis of generalized linear models (GLMs). A graphical model is associated with a graph $ \mathcal{G} = (\mathcal{V}, \mathcal{E}) $, where the node set $ \mathcal{V}$ represents the variables of interest, and the edge set $ \mathcal{E} \subseteq \mathcal{V} \times \mathcal{V} $ encodes the conditional dependence relationships. The goal of a graphical model is to accurately estimate $\mathcal{E}$ for a given set of $\mathcal{V}$.

\subsection{Modeling}
\label{sec:Modeling}

To simplify notation, we consider the scenario with a continuous and zero inflated clinical outcome (i.e., medical costs). It is noted that the model can be readily extended to accommodate multiple outcomes with distributions belonging to the exponential family.\\

In what follows, we use the uppercase letters to denote random variables, and the lowercase letters to denote observed values. Let $ Y_j^{(t)} $ be the outcome of disease $ j $ at time $ t $, $j = 1,\ldots,d$ where $d$ is the number of disease conditions, and $t = 1,\ldots,T$ where $T$ is number of time periods. To accommodate zero-inflation, we introduce a pseudo outcome $ V_{j}^{(t)} $ indicating the presence of treatment for disease $ j $ at time $ t $. Accordingly, $\boldsymbol{Y}^{(t)} = (Y_{1}^{(t)}, \ldots, Y_{d}^{(t)})' $ and $\boldsymbol{V}^{(t)} = (V_{1}^{(t)}, \ldots, V_{d}^{(t)})' $ denote outcome and pseudo outcome of all diseases at time $ t $, respectively. And we use $\boldsymbol{Y}_{-j}^{(t)}$ and $\boldsymbol{V}_{-j}^{(t)}$ to represent counterparts of $\boldsymbol{Y}^{(t)}$ and $\boldsymbol{V}^{(t)}$ with the components corresponding to disease $ j $ removed. \\

Motivated by a two-part GLM model \citep{mei2023human} for zero-inflated data, we consider the following model for disease $j$ at time $t$:

\begin{equation}
    \label{equ:ref_model}
    \begin{cases}
        p(V_{j}^{(t)}=v_{j}^{(t)}|\boldsymbol{v}_{-j}^{(t)},\boldsymbol{y}_{-j}^{(t)})=C_{0}(v_{j}^{(t)})\exp[\vartheta_{0}v_{j}^{(t)}-b_{0}(\vartheta_{0})],\quad v_{j}^{(t)}\in\{0,1\},\\
        p(Y_{j}^{(t)}=y_{j}^{(t)}|V_{j}^{(t)}=0,\boldsymbol{v}_{-j}^{(t)},\boldsymbol{y}_{-j}^{(t)})=I\{y_j^{(t)}=0\},\\
        p(Y_{j}^{(t)}=y_{j}^{(t)}|V_{j}^{(t)}=1,\boldsymbol{v}_{-j}^{(t)},\boldsymbol{y}_{-j}^{(t)})=C_{1}(y_{j}^{(t)})\exp[\vartheta_{1}y_{j}^{(t)}-b_{1}(\vartheta_{1})],\quad y_j^{(t)}>0,\\
    \end{cases}
\end{equation}

where $\{C_0,C_1,\vartheta_{0},\vartheta_{1}\}$ and $\{b_0,b_1\}$ are parameters and functions defined in standard GLMs.
Use $\mu_{j}^{(t)}$ and $p_j^{(t)}$ to denote the conditional expectation of $Y_{j}^{(t)}$ and $V_{j}^{(t)}$, respectively. We have

\begin{equation}
    \begin{aligned}
        p_{j}^{(t)}&=E(V_{j}^{(t)}|\boldsymbol{v}_{-j}^{(t)},\boldsymbol{y}_{-j}^{(t)}),\\
        \mu_{j}^{(t)}&=E(Y_{j}^{(t)}|V_{j}^{(t)}=1,\boldsymbol{v}_{-j}^{(t)},\boldsymbol{y}_{-j}^{(t)}),\\
    \end{aligned}
\end{equation}

In our data analysis, $V_{j}^{(t)}$ is the indicator of treatment and $Y_{j}^{(t)}$ is the cost of disease $j$ at time $t$. Based on the previously defined model (\ref{equ:ref_model}), we further assume


\begin{equation}
    \begin{aligned}
        &f(V_j^{(t)}=v_j^{(t)}|\boldsymbol{v}_{-j}^{(t)},\boldsymbol{y}_{-j}^{(t)})=\exp[v_j^{(t)}\log(\frac{p_j^{(t)}}{1-p_j^{(t)}})+\log(1-p_j^{(t)})],\\
        &f(Y_{j}^{(t)}=y_{j}^{(t)}|V_{j}^{(t)}=1,\boldsymbol{v}_{-j}^{(t)},\boldsymbol{y}_{-j}^{(t)})=\exp[\frac{(y_{j}^{(t)}\mu_j^{(t)}-\frac{{\mu_j^{(t)}}^2}{2})}{{\sigma_j^{(t)}}^2}-\frac{1}{2}(\frac{{y_{j}^{(t)}}^2}{{\sigma_j^{(t)}}^2}+\log(2\pi{\sigma_j^{(t)}}^2))],
    \end{aligned}
    \label{equ:distribution}
\end{equation}

with the logit link function $\log(\frac{p_j^{(t)}}{1-p_j^{(t)}}) = \zeta_{j}^{(t)}+r_{j}^{(t)}$ and the natural link function $\mu_{j}^{(t)}=\alpha_{j}^{(t)}+\tau_{j}^{(t)}r_{j}^{(t)}$, where $r_{j}^{(t)}={\boldsymbol{v}_{-j}^{(t)}}'\boldsymbol{\eta}_{j}^{(t)} + {\boldsymbol{y}_{-j}^{(t)}}'\boldsymbol{\beta}_{j}^{(t)}$ represents the regression part in GLMs and $\boldsymbol{\eta}_{j}^{(t)}, \boldsymbol{\beta}_{j}^{(t)}$ denote regression coefficients.
Consequently, the density function of $(V_j^{(t)},Y_j^{(t)})$ given $(\boldsymbol{V}_{-j}^{(t)},\boldsymbol{Y}_{-j}^{(t)})$ can be derived as

\begin{equation}
    \begin{aligned}
        f(V_{j}^{(t)}=v_j^{(t)},Y_j^{(t)}=y_j^{(t)}|\boldsymbol{v}_{-j}^{(t)},\boldsymbol{y}_{-j}^{(t)})&=f(V_j^{(t)}=v_j^{(t)}|\boldsymbol{v}_{-j}^{(t)},\boldsymbol{y}_{-j}^{(t)})f(Y_j^{(t)}=y_j^{(t)}|V_j^{(t)}=v_j^{(t)},\boldsymbol{v}_{-j}^{(t)},\boldsymbol{y}_{-j}^{(t)})\\
        &=({\frac{e^{\zeta_{j}^{(t)}+r_j^{(t)}}}{1+e^{\zeta_{j}^{(t)}+r_j^{(t)}}}})^{v_{j}^{(t)}}({\frac{1}{1+e^{\zeta_{j}^{(t)}+r_j^{(t)}}}})^{1-v_{j}^{(t)}}\cdot\\
        &\quad (\frac{1}{\sqrt{2\pi {\sigma_j^{(t)}}^{2}}}e^{-\frac{(y_j^{(t)}-\alpha_j^{(t)}-\tau_j^{(t)} r_j^{(t)})^2}{2{\sigma_j^{(t)}}^{2}}})^{v_{j}^{(t)}}(\delta_0(y_j^{(t)}))^{1-v_j^{(t)}},
    \end{aligned}
    \label{equ:density}
\end{equation}

where $\delta_0(y_j^{(t)})$ is the dirac delta function.
It's noted that $(\delta_0(y_j^{(t)}))^{1-v_j^{(t)}}$ is not correlated to the unknown parameters.
Hence, we can write the log-likelihood for disease $j$ at time $t$ as

\begin{equation}
    \begin{aligned}
        &l_j^{(t)}(\boldsymbol{\alpha}_{j}^{(t)},\boldsymbol{\zeta}_j^{(t)},\boldsymbol{\tau}_j^{(t)},\sigma_j^{(t)},\boldsymbol{\eta}_j^{(t)},\boldsymbol{\beta}_j^{(t)}|\boldsymbol{Y}_{-j}^{(t)}=\boldsymbol{y}_{-j}^{(t)},\boldsymbol{V}_{-j}^{(t)}=\boldsymbol{v}_{-j}^{(t)})\\
        =&-\frac{1}{n}\sum_{i=1}^n\big\{\log[1+\exp(\zeta_j^{(t)}+r_{ij}^{(t)})]\\
        &+\sum_{i=1}^nv_{ij}^{(t)}[\zeta_{j}^{(t)}+r_{ij}^{(t)}-\frac{1}{2{\sigma_j^{(t)}}^{2}}(y_{ij}^{(t)}
        -\alpha_{j}^{(t)}-\tau_{j}^{(t)}r_{ij}^{(t)})^2
        -\frac{1}{2}(\log{{\sigma_j^{(t)}}^{2}}+\log2\pi)]\big\},
    \end{aligned}
    \label{equ:log_likelihood}
\end{equation}

where \( n \) represents the sample size.

\subsection{Estimation}
\label{sec:Estimation}

In the literature, there are multiple approaches for estimating the structure of a graph. Motivated by the approach of neighborhood selection, \citep{peng2021reinforced} we establish the idea that understanding the connectivity of individual nodes at each time period allows determination of the entire graph's time-varying structure. The objective is to estimate $\boldsymbol{\beta}_j^{(t)}$ and $\boldsymbol{\eta}_j^{(t)}$ with the log-likelihood functions defined for individual nodes and time points (\ref{equ:log_likelihood}). 
We suppose that two nodes are conditionally independent if and only if all the related parameters are zero. Denote the parameters connecting nodes $j$ and $m$ at time $t$ as $\boldsymbol{\theta}_{j,m}^{(t)}=(\beta_{j,m}^{(t)},\eta_{j,m}^{(t)})$, where $m\in \boldsymbol{M}=\{1,...,d\}\backslash\{j\}$. $||\theta_{j,m}^{(t)}||_2 = 0$ indicates that nodes $j$ and $m$ at time $t$ are conditionally independent. 
Therefore, we adopt a penalty term $p_1$ for group variable selection.
In addition, to infer a time-dependent sequence of networks, we introduce a another penalty term $p_2$ encouraging $\boldsymbol{\theta}_j^{(t)}=(\theta_{j,m}^{(t)})_{m\in \boldsymbol{M}}$ to smoothly vary over time. Our objective function is

\begin{equation}
    \begin{aligned}
        \underset{\boldsymbol{\alpha}_{j}^{(t)},\boldsymbol{\zeta}_j^{(t)},\boldsymbol{\tau}_j^{(t)},\sigma_j^{(t)},\boldsymbol{\theta}_j^{(t)}}{\min}& -\sum_{t=1}^T l_j^{(t)}(\boldsymbol{\alpha}_{j}^{(t)},\boldsymbol{\zeta}_j^{(t)},\boldsymbol{\tau}_j^{(t)},\sigma_j^{(t)},\boldsymbol{\theta}_j^{(t)}|\boldsymbol{Y}_{-j}^{(t)}=\boldsymbol{y}_{-j}^{(t)},\boldsymbol{V}_{-j}^{(t)}=\boldsymbol{v}_{-j}^{(t)})\\
        &+\sum_{t=1}^{T}\sum_{m\neq j}p_1(\theta_{j,m}^{(t)};\lambda_1) + \sum_{t=2}^{T}p_2(\boldsymbol{\theta}_{j}^{(t)}-\boldsymbol{\theta}_{j}^{(t-1)};\lambda_2).
    \end{aligned}
    \label{equ:obj}
\end{equation}

In the objective function, the first term represents the sum of the log-likelihood function at all time periods. The second term introduces a penalty to promote network sparsity. This is motivated by the observation from Figure \ref{fig:heatmap} that not all diseases are interconnected. The third term encourages similarity between consecutive time periods. 
Alternative methods exist for conducting time-dependent analyses, such as constructing the objective function as a function of time $t$. \citep{rahili2016distributed,danieli2020modeling} Given that medical costs are typically aggregated within a fixed time window, our chosen approach is favored over methods involving continuously varying $t$. 
By aggregating medical costs annually, we are limited to data at discrete, designated time points. This aggregation overlooks information on cost variations that occur within the intervals between these points, such as seasonal fluctuations. Consequently, modeling costs as a smooth and differentiable function of time is not appropriate. While it is possible to aggregate costs over shorter intervals, such as monthly, doing so would compromise the statistical power, increase computational demands, and diminish the interpretability of the results. Moreover, our analysis of temporal variations primarily focuses on long-term trends rather than short-term seasonal fluctuations.
 

\subsection{Statistical properties}


To accommodate the nature of medical costs data, we assume in Section 2.1 that $Y_{j}^{(t)}|V_{j}^{(t)}=1$ follows a Gaussian distribution. However, it is important to note that within the GLM framework, the model is easily extendable to accommodate other types of outcomes from the exponential family. In this section, we present the consistency of estimation and variable selection, considering a broader definition of the density function $f$. While identifiability of 
a general form of $f$ is typically imposed as an assumption, we specifically prove it under the Gaussian assumption. For other exponential family outcomes, identifiability can be established in a similar manner.\\

Let {$\boldsymbol{\gamma}_j^{(t)}=(\boldsymbol{\alpha}_{j}^{(t)},\boldsymbol{\zeta}_j^{(t)},\boldsymbol{\tau}_j^{(t)},\sigma_j^{(t)},\boldsymbol{\theta}_j^{(t)})$} denote the vector of unknown parameters for disease $j$ at time $t$ and let the vector $\boldsymbol{\Phi}_j=(\boldsymbol{\gamma}_j^{(1)},...,\boldsymbol{\gamma}_j^{(T)})'$ in $\mathbb{R}^{(2d+2)T \times 1}$ denote all unknown parameters for disease $j$.
Rewrite $\gamma_j^{(t)}=
(\gamma_{j1}^{(t)},\gamma_{j2}^{(t)})$, where $\gamma_{j1}^{(t)}$ consists of all the nonzero parameters (i.e., $(\boldsymbol{\alpha}_{j}^{(t)},\boldsymbol{\zeta}_j^{(t)},\boldsymbol{\tau}_j^{(t)},\sigma_j^{(t)})$ and the nonzero regression coefficients in $\boldsymbol{\theta}_{j}^{(t)}$) and $\gamma_{j2}^{(t)}$ contains all the zero regression coefficients in $\boldsymbol{\theta}_{j}^{(t)}$.
Accordingly, let \(\boldsymbol{\Phi}_{j1}=(\gamma_{j1}^{(1)},...,\gamma_{j1}^{(T)})'\) denote  the set of non-zero parameters and \(\boldsymbol{\Phi}_{j2}\) denote the set of zero regression coefficients.
To simplify notation, we define $l_j(\boldsymbol{\gamma}_j^{(t)})=nl_j^{(t)}(\boldsymbol{\alpha}_{j}^{(t)},\boldsymbol{\zeta}_j^{(t)},\boldsymbol{\tau}_j^{(t)},\sigma_j^{(t)},\boldsymbol{\theta}_j^{(t)}|\boldsymbol{y}_{-j}^{(t)},\boldsymbol{v}_{-j}^{(t)})$.
Let $\hat{\boldsymbol{\Phi}}_{j}=(\hat{\boldsymbol{\Phi}}_{j1},\hat{\boldsymbol{\Phi}}_{j2})$ denote the minimizer of 

\begin{equation}
    \begin{aligned}
        Q_n(\boldsymbol{\Phi}_j)=-\sum_{t=1}^T l_j(\boldsymbol{\gamma}_j^{(t)})+n\sum_{t=1}^{T}\sum_{m\neq j}p_1(\theta_{j,m}^{(t)};\lambda_1) + n\sum_{t=2}^{T} p_2(\boldsymbol{\theta}_{j}^{(t)}-\boldsymbol{\theta}_{j}^{(t-1)};\lambda_2),
    \end{aligned}
    \label{equ:objective}
\end{equation}

where $p_1(\cdot;\lambda_1)$ and $p_2(\cdot;\lambda_2)$ denote two penalty functions.
To derive theoretical results, we introduce several widely used assumptions.

\begin{assumption}
    \label{asmpt:finite}
    The number of time periods $T$ and the number of nodes $d$ satisfy
    \begin{equation}
        T < \infty \quad \text{and} \quad d < \infty \quad \text{as} \quad n \to \infty.
    \end{equation}
\end{assumption}

    

\begin{assumption}
    \label{asmpt:identifiability}
    For any disease $j$ and time $t$ , the density function $f(v_{j}^{(t)},y_{j}^{(t)};\boldsymbol{\gamma}_{j}^{(t)})$ is identifiable, meaning that for any $\boldsymbol{\gamma}_j^{(t)}$ and $ \boldsymbol{\tilde{\gamma}}_j^{(t)}$ such that $f(v_j^{(t)},y_j^{(t)};\boldsymbol{\gamma}_j^{(t)}) = f(v_j^{(t)},y_j^{(t)};\boldsymbol{\tilde{\gamma}}_j^{(t)})$ (with probability 1), we have $\boldsymbol{\gamma}_j^{(t)} = \boldsymbol{\tilde{\gamma}}_j^{(t)}$.
\end{assumption}

\begin{assumption}
    \label{asmpt:condition 1}
    For any disease $j$ and time $t$ , the density function $f(v_{j}^{(t)},y_{j}^{(t)};\boldsymbol{\gamma}_{j}^{(t)})$ has a common support. Moreoever, it satisfies:
    \begin{equation}
        \begin{aligned}
            &E\big[\frac{\partial\log f(v_{j}^{(t)},y_{j}^{(t)};\boldsymbol{\gamma}_{j}^{(t)})}{\partial \boldsymbol{\gamma}}\Big|_{\boldsymbol{\gamma}=\boldsymbol{\gamma}_{j,0}^{(t)}} \big] = 0,\\
            E\big[\frac{\partial\log f(v_{j}^{(t)},y_{j}^{(t)};\boldsymbol{\gamma}_{j}^{(t)})}{\partial \boldsymbol{\gamma}_{j,k}^{(t)}}&\frac{\partial\log  f(v_{j}^{(t)},y_{j}^{(t)};\boldsymbol{\gamma}_{j}^{(t)})}{\partial \boldsymbol{\gamma}_{j,l}^{(t)}} \big] = 
            E\big[-\frac{\partial^2\log f(v_{j}^{(t)},y_{j}^{(t)};\boldsymbol{\gamma}_{j}^{(t)})}{\partial \boldsymbol{\gamma}_{j,k}^{(t)}\partial \boldsymbol{\gamma}_{j,l}^{(t)}} \big],
        \end{aligned}
    \end{equation}
    where $\boldsymbol{\gamma}_{j,0}^{(t)}$ represents the true parameters and $\boldsymbol{\gamma}_{j,k}^{(t)}$ and $\boldsymbol{\gamma}_{j,l}^{(t)}$ represent the $k$-th and $l$-th elements of $\boldsymbol{\gamma}_{j}^{(t)}$, respectively.
\end{assumption}

\begin{assumption}
    \label{asmpt:condition 2}
    The Fisher information matrix for $\boldsymbol{\gamma}_{j}^{(t)}$
    
    \begin{equation}
        \begin{aligned}
            \mathcal{I}(\boldsymbol{\gamma}_{j}^{(t)})=E\Big\{\big[\frac{\partial \log f(v_j^{(t)},y_j^{(t)};\boldsymbol{\gamma}_{j}^{(t)})}{\partial \boldsymbol{\gamma}_{j}^{(t)}}\big]\big[\frac{\partial \log f(v_j^{(t)},y_j^{(t)};\boldsymbol{\gamma}_{j}^{(t)})}{\partial \boldsymbol{\gamma}_{j}^{(t)}}\big]'\Big\},
        \end{aligned}
    \end{equation}
    
    is finite and positive-definite at $\boldsymbol{\gamma}_{j}^{(t)}=\boldsymbol{\gamma}_{j,0}^{(t)}$.
\end{assumption}

\begin{assumption}
    \label{asmpt:condition 3}
    There exists an open set $\mathcal{N}_0$ that contains the true parameters $\boldsymbol{\gamma}_{j,0}^{(t)}$, such that the density $f(v_{j}^{(t)},y_{j}^{(t)};\boldsymbol{\gamma}_{j}^{(t)})$ admits all third 
    derivatives for all $\boldsymbol{\gamma}_{j}^{(t)}\in \mathcal{N}_0$. There exist integrable functions $B_{h,k,l}(v_{j}^{(t)},y_{j}^{(t)})$ for all $h,j,l$ such that 
    
    \begin{equation}
        \begin{aligned}
            \Big|\frac{\partial^3}{\partial\boldsymbol{\gamma}_{j,h}^{(t)}\partial\boldsymbol{\gamma}_{j,k}^{(t)}\partial\boldsymbol{\gamma}_{j,l}^{(t)}}\log f(v_{j}^{(t)},y_{j}^{(t)};\boldsymbol{\gamma}_{j}^{(t)})\Big| \leq B_{h,k,l}(v_{j}^{(t)},y_{j}^{(t)}),
        \end{aligned}
    \end{equation}
    
    where $E[B_{h,k,l}({v_{j}^{(t)}},{y_{j}^{(t)}})] < \infty $ and $\boldsymbol{\gamma}_{j,h}^{(t)}$ is the $h$-th element of $\boldsymbol{\gamma}_{j}^{(t)}$.
\end{assumption}

\begin{assumption}
    \label{asmpt:condition 4}
    $\sqrt{n}\min\{\lambda_1,\lambda_2\}\rightarrow\infty, \underset{n\rightarrow\infty}{\liminf}\underset{c\rightarrow 0^+}{\liminf}p^\prime_1(c;\lambda_1)/\lambda_1>0,\underset{n\rightarrow\infty}{\liminf}\underset{c\rightarrow 0^+}{\liminf}p^\prime_2(c;\lambda_2)/\lambda_2>0$.
\end{assumption}

Assumption \ref{asmpt:finite} asserts that the number of time periods $T$ and the number of nodes $d$ remain finite as the sample size increases.
Assumptions \ref{asmpt:identifiability}-\ref{asmpt:condition 4} are standard regularity conditions commonly used in the literature to derive asymptotic results. \citep{meinshausen2006high, yuan2007model, fan2004nonconcave}
Among them, Assumption \ref{asmpt:identifiability} assumes identifiability of a general form of density function $f$. 
Assumptions \ref{asmpt:condition 1}-\ref{asmpt:condition 3} impose constraints on the second and fourth moments of the likelihood function. The information matrix is assumed to be positive definite.
Assumption \ref{asmpt:condition 4} sets constraints on tuning parameters $\lambda_1$ and $\lambda_2$ with respect to the sample size, ensuring that the penalty terms allow for the distinction between non-zero and zero coefficients.

\begin{theorem}
    Suppose $(\boldsymbol{V}_j^{(t)},\boldsymbol{Y}_j^{(t)})$ follows the previously defined density (\ref{equ:density}).
    Then, for any $\boldsymbol{\gamma}_j^{(t)}$ and ${\boldsymbol{\tilde{\gamma}}_j^{(t)}}$ such that $f(v_j^{(t)},y_j^{(t)};\boldsymbol{\gamma}_j^{(t)}) = f(v_j^{(t)},y_j^{(t)};\boldsymbol{\tilde{\gamma}}_j^{(t)})$ (with probability 1), we have $\boldsymbol{\gamma}_j^{(t)} = \boldsymbol{\tilde{\gamma}}_j^{(t)}$.
    \label{thm:identifiablility}
\end{theorem}

\begin{theorem}
    Under Assumptions \ref{asmpt:finite}-\ref{asmpt:condition 3}, there exists a local minimizer $\hat{\boldsymbol{\Phi}}_{j}$ of $Q_n(\boldsymbol{\Phi}_j)$ such that
    \begin{equation*}
        ||\hat{\boldsymbol{\Phi}}_j-\boldsymbol{\Phi}_{j,0}||=O_p(n^{-\frac{1}{2}}+a_n),
    \end{equation*}
    where $a_n=\max\Big\{\underset{m,t}{\max}\{|p_1^{\prime}(\theta_{(j,m),0}^{(t)};\lambda_1)|;\theta_{(j,m),0}^{(t)}\neq0\},\underset{t}{\max}\{|p_2^{\prime}(\boldsymbol{\theta}_{j,0}^{(t)}-\boldsymbol{\theta}_{j,0}^{(t-1)};\lambda_2)|;\boldsymbol{\theta_{j,0}^{(t)}}-\boldsymbol{\theta_{j,0}^{(t-1)}}\neq\boldsymbol{0} \} \Big\}$. Here, $\theta_{(j,m),0}^{(t)},\boldsymbol{\theta_{j,0}^{(t)}}$ and $\boldsymbol{\Phi}_{j,0}$ denote the true parameters.
    \label{thm:consistency}
\end{theorem}

\begin{theorem}
    Under Assumptions \ref{asmpt:finite}-\ref{asmpt:condition 4}, if $a_n=O(n^{-\frac{1}{2}})$, $\hat{\boldsymbol{\Phi}}_{j}=(\hat{\boldsymbol{\Phi}}_{j1},\hat{\boldsymbol{\Phi}}_{j2})$ achives variable selection consistency, meaning that $||\hat{\boldsymbol{\Phi}}_{j1}-\boldsymbol{\Phi}_{j1,0}||=O_p(n^{-\frac{1}{2}}+a_n)$ and $P(\hat{\boldsymbol{\Phi}}_{j2}=\boldsymbol{0}) \rightarrow 1$.
    \label{thm:model selection consistency}
\end{theorem}

Theorem \ref{thm:identifiablility} demonstrates the identifiability of parameters in the proposed model under the Gaussian assumption. For a general form of density function $f$ assuming identifiability, we further establish the convergence rate of the Maximum Likelihood Estimator for fixed $(T, d)$ as $n\rightarrow \infty$.
By Theorem \ref{thm:consistency}, appropriate choices of the tuning parameters $\lambda_1, \lambda_2$ lead to a $\sqrt{n}$-consistency rate estimator.
Since it does not guarantee zero estimates for the true zero coefficients, we introduce Theorem \ref{thm:model selection consistency} for selection consistency of the regression coefficients. A wide ranges of commonly used penalty forms can 
satisfy the established properties, including Smoothly Clipped Absolute Deviation, \citep{fan2001variable} Minimax Concave Penalty, \citep{zhang2010nearly} and adaptive LASSO. \citep{zou2006adaptive} However, it should be noticed that the LASSO penalty is not included. For the $\ell_1$ penalty, $a_n=\lambda_n$, so Theorem \ref{thm:consistency} requires that $\lambda_n=O_p(n^{-1/2})$. Conversely, Theorem \ref{thm:model selection consistency} requires that $\sqrt{n}\lambda_n \rightarrow\infty$.
Detailed proofs can be found in Section B of the Supplementary Materials.
While considerations of conditional independence, group selection, and smoothed temporal variations add complexity of the proof, developments of Theorem 1-3 for coefficients validates the proposed model and the estimation approach.

\subsection{Computational algorithm}
\label{sec:computate}

To address the objective function (\ref{equ:obj}), we adopt the adaptive group LASSO for group variable selection and the adaptive fused LASSO for temporal smoothness.
Since the objective function contains convex penalized terms,
we employ an appropriate solver known as Alternating Direction Method of Multipliers (ADMM). \citep{boyd2011distributed} This solver allows us to decouple the variables, resulting in a separable minimization problem that can be efficiently solved in parallel. 
The sub-problems within this iterative algorithm utilize proximal operators, which can  be resolved using closed-form solutions. To transform the original problem into a separable form,  we introduce a consensus variable $\boldsymbol{Z}=\{\boldsymbol{Z}_2,...,\boldsymbol{Z}_T\}$, then rewrite the problem as

\begin{equation}
    \begin{aligned}
        \underset{\boldsymbol{\alpha}_{j}^{(t)},\boldsymbol{\zeta}_j^{(t)},\boldsymbol{\tau}_j^{(t)},\sigma_j^{(t)},\boldsymbol{\theta}_j^{(t)}}{\min}& -\sum_{t=1}^T l_j^{(t)}(\boldsymbol{\alpha}_{j}^{(t)},\boldsymbol{\zeta}_j^{(t)},\boldsymbol{\tau}_j^{(t)},\sigma_j^{(t)},\boldsymbol{\theta}_j^{(t)}|\boldsymbol{Y}_{-j}^{(t)}=\boldsymbol{y}_{-j}^{(t)},\boldsymbol{V}_{-j}^{(t)}=\boldsymbol{v}_{-j}^{(t)}),\\
        &+\lambda_1\sum_{t=1}^{T}\sum_{m\neq j}\omega_{1,m}^{(t)}||\theta_{j,m}^{(t)}||_2+ \lambda_2\sum_{t=2}^{T}\omega_{2}^{(t)}||\boldsymbol{\theta}_{j}^{(t)}-\boldsymbol{\theta}_{j}^{(t-1)}||_1\\
        &{\rm s.t.}\quad\quad\quad\quad\quad \boldsymbol{\theta}_{j}^{(t)}-\boldsymbol{\theta}_{j}^{(t-1)}=\boldsymbol{Z}_t,\quad t=2,...,T, 
    \end{aligned}
\end{equation}

here $\omega_{1,m}^{(t)}$ and $\omega_{2}^{(t)}$ are the adaptive weights. We set $\omega_{1,m}^{(t)}=||{\theta_{j,m}^{(t)}}^{0}||_2^{-1}$ and $\omega_{2}^{(t)}=||{\boldsymbol{\theta}_{j}^{(t)}}^{0}-{\boldsymbol{\theta}_{j}^{(t-1)}}^0||_1^{-1}$ in simulation study and real data analysis, where ${\theta_{j,m}^{(t)}}^{0},{\boldsymbol{\theta}_{j}^{(t)}}^{0}$ denote the initial values. The corresponding augmented Lagrangian is expressed as

\begin{equation}
    \label{equ:lagrangian}
    \begin{aligned}
        \mathcal{L}_{\rho}(\boldsymbol{\Phi}_j,\boldsymbol{Z},\boldsymbol{U})=&-\sum_{t=1}^T l_j^{(t)}(\boldsymbol{\alpha}_{j}^{(t)},\boldsymbol{\zeta}_j^{(t)},\boldsymbol{\tau}_j^{(t)},\sigma_j^{(t)},\boldsymbol{\theta}_j^{(t)}|\boldsymbol{Y}_{-j}^{(t)}=\boldsymbol{y}_{-j}^{(t)},\boldsymbol{V}_{-j}^{(t)}=\boldsymbol{v}_{-j}^{(t)})+\lambda_1\sum_{t=1}^{T}\sum_{m\neq j}\omega_{1,m}^{(t)}||\theta_{j,m}^{(t)}||_2 \\
        &+\lambda_2\sum_{t=2}^{T}\omega_{2}^{(t)}||\boldsymbol{Z}_t||_1 +\frac{\rho}{2}\sum_{t=2}^T(||\boldsymbol{\theta}_j^{(t)}-\boldsymbol{\theta}_j^{(t-1)}-\boldsymbol{Z}_t+\boldsymbol{U}_t||_2^2-||\boldsymbol{U}_t||_2^2),
    \end{aligned}
\end{equation}

where $\boldsymbol{U}=\{\boldsymbol{U}_2,...,\boldsymbol{U}_T\}$ contains the scaled dual variables and $\rho>0$ is the augmented Lagrangian parameter. 
Denote $k$ as the iteration number, each updating step of ADMM is shown as

\begin{equation}
    \label{equ:iter}
    \begin{aligned}
        \boldsymbol{\Phi}_j^{k+1}&=\underset{\boldsymbol{\Phi_j}}{\mathop{\arg\min}}\ \mathcal{L}_{\rho}(\boldsymbol{\Phi}_j,\boldsymbol{Z}^k,\boldsymbol{U}^k),\\
        \boldsymbol{Z}^{k+1}&=\underset{\boldsymbol{Z}}{\mathop{\arg\min}}\ \mathcal{L}_{\rho}(\boldsymbol{\Phi}_j^{k+1},\boldsymbol{Z},\boldsymbol{U}^k),\\
        \boldsymbol{U}^{k+1}&=\underset{\boldsymbol{U}}{\mathop{\arg\min}}\ \mathcal{L}_{\rho}(\boldsymbol{\Phi}_j^{k+1},\boldsymbol{Z}^{k+1},\boldsymbol{U}).
    \end{aligned}
\end{equation}

(a) $\boldsymbol{\Phi}_j$-update\\

Regarding the component involving $\boldsymbol{\Phi}_j$ in (\ref{equ:lagrangian}),
the objective function can be expressed as the sum of a differentiable function $g(\boldsymbol{\Phi}_j)=-\sum_{t=1}^T l_j^{(t)}+\frac{\rho}{2}\sum_{t=2}^T(||\boldsymbol{\theta}_j^{(t)}-\boldsymbol{\theta}_j^{(t-1)}-\boldsymbol{Z}_t+\boldsymbol{U}_t||_2^2-||\boldsymbol{U}_t||_2^2)$ 
and a nondifferentiable convex function $h(\boldsymbol{\Phi}_j)=\lambda_1\sum_{t=1}^T\sum_{m\neq j}\omega_{1,m}^{(t)}||\theta_{j,m}^{(t)}||_2$.
Thus, to obtain $\boldsymbol{\Phi}_j^{k+1}$ in each updating step of ADMM, we adopt the proximal gradient descent algorithm for optimization.
Specifically, the proximity operator for $\boldsymbol{\Phi_j}$ is 

\begin{equation}
    {\rm \textbf{prox}}_{h,a}(\boldsymbol{\Phi}_j)=\underset{\boldsymbol{x}}{\arg\min}\frac{1}{2a}||\boldsymbol{\Phi}_j-\boldsymbol{x}||_2^2+h(\boldsymbol{\Phi}_j),
\end{equation}

where {$a$} is the step size. With a set of initial values, the update rule can be described as

\begin{equation}
    \boldsymbol{\Phi}_j^{k+1}={\rm \textbf{prox}}_{h,a_k}(\boldsymbol{\Phi}_{j}^k-a_k\nabla g(\boldsymbol{\Phi}_j^k)),
\end{equation}

where $a_k$ is the step size of $k$-th iteration. The iterative algorithm terminates when $|\boldsymbol{\Phi}_j^{k+1}-\boldsymbol{\Phi}_j^{k}|<\epsilon$ with $\epsilon$ being pre-specified.
As $h(\boldsymbol{\Phi}_j)$ only involves $\boldsymbol{\theta}_j^{(t)}$, the gradient descent algorithm is used for updating $\boldsymbol{\alpha}_j^{(t)},\boldsymbol{\zeta}_j^{(t)},\boldsymbol{\tau}_j^{(t)},\sigma_j^{(t)}$:

\begin{equation}
    \begin{aligned}
        {\boldsymbol{\alpha}_j^{(t)}}^{k+1}={\boldsymbol{\alpha}_j^{(t)}}^{k}-a_k\partial g({\boldsymbol{\alpha}_j^{(t)}}^{k}), \quad t=1,...,T,\\
        {\boldsymbol{\zeta}_j^{(t)}}^{k+1}={\boldsymbol{\zeta}_j^{(t)}}^{k}-a_k\partial g({\boldsymbol{\zeta}_j^{(t)}}^{k}), \quad t=1,...,T,\\
        {\boldsymbol{\tau}_j^{(t)}}^{k+1}={\boldsymbol{\tau}_j^{(t)}}^{k}-a_k\partial g({\boldsymbol{\tau}_j^{(t)}}^{k}), \quad t=1,...,T,\\
        {\sigma_j^{(t)}}^{k+1}={\sigma_j^{(t)}}^{k}-a_k\partial g({\sigma_j^{(t)}}^{k}), \quad t=1,...,T.
    \end{aligned}
\end{equation}

Then, we set $\phi({\theta_{j,m}^{(t)}}^k)={\theta_{j,m}^{(t)}}^{k}-a_k\partial g({\theta_{j,m}^{(t)}}^{k})$, the update rule for ${\theta_{j,m}^{(t)}}(m\neq j)$ is 

\begin{equation}
    \begin{aligned}
        {\theta_{j,m}^{(t)}}^{k+1}=\begin{cases}
        0\quad \quad \quad \quad \quad \quad \quad \quad \quad \quad\quad\quad\ \ \ ,\ {\rm if}||\phi({\theta_{j,m}^{(t)}}^{k})||\leq a_k\lambda_1\omega_{1,m}^{(t)},\\
        \frac{(||\phi({\theta_{j,m}^{(t)}}^{k})-a_k\lambda_1\omega_{1,m}^{(t)}||))\phi({\theta_{j,m}^{(t)}}^{k})}{||\phi({\theta_{j,m}^{(t)}}^{k})||},\ {\rm if}||\phi({\theta_{j,m}^{(t)}}^{k})||>a_k\lambda_1\omega_{1,m}^{(t)}.
        \end{cases}
    \end{aligned}
\end{equation}

(b) $\boldsymbol{Z}$-update

$\boldsymbol{Z}$ can be written as the proximal operator of the $\ell_1$-norm, which has the closed-form solution:

\begin{equation}
    \label{equ:Z update}
    \boldsymbol{Z}_t^{k+1}={\rm \textbf{prox}}_{||\cdot||_1,\frac{\lambda_2\omega_2^{(t)}}{\rho}}({\boldsymbol{\theta}_j^{(t)}}^{k+1}-{\boldsymbol{\theta}_j^{(t-1)}}^{k+1}+\boldsymbol{U}_t^{k}),\quad t=2,...,T.
\end{equation}

(c) $\boldsymbol{U}$-update

The update of a scaled dual variable can be easily derived as

\begin{equation}
    \label{equ:U update}
    \boldsymbol{U}_t^{k+1}=\boldsymbol{U}_t^{k}+({\boldsymbol{\theta}_j^{(t)}}^{k+1}-{\boldsymbol{\theta}_j^{(t-1)}}^{k+1}-\boldsymbol{Z}_t^{k+1}),\quad {t}=2,...,T.
\end{equation}

In summary, we split the problem into a series of sub-problems 
and use iterative optimization algorithms to get the optimal solution.
Computation can be conducted in parallel for each node and each time period to reduce computational time.
Existing works have shown that this ADMM-based algorithm guarantees convergence, \citep{boyd2011distributed, nishihara2015general} which is also confirmed in our simulation studies. 
Choosing tuning parameters $\lambda_1$ and $\lambda_2$ through cross-validation, Algorithm \ref{alg:alg1} shows the detailed procedure and the R codes are available at \textit{github.com/haomei/Dynamic\_HDN\_Costs}.

\begin{algorithm}[htbp]
    \caption{Find the optimal parameter $\boldsymbol{\theta}^{(1)},...,\boldsymbol{\theta}^{(T)}$ for the problem}
    \label{alg:alg1}
    \renewcommand{\algorithmicrequire}{\textbf{Input:}}
    \renewcommand{\algorithmicensure}{\textbf{Output:}}
    \begin{algorithmic}
        \Require \ \\
        Dataset $\boldsymbol{Y^{(1)},V^{(1)}},...,\boldsymbol{Y^{(T)},V^{(T)}}$;\\
        Tuning parameter $\lambda_1,\lambda_2,\rho$;\\
        Threshold $\epsilon$;\\
        
        \Ensure \ \\      
        $\boldsymbol{\hat{\theta}}_{j}^{(1)},...,\boldsymbol{\hat{\theta}_{j}}^{(T)}, j=1,...,d$ ;\\     
        \ \\ 
        \setcounter{ALG@line}{0} 
        \For {j=1,..,d}
            \State k = 0;
            \State Initialize $\boldsymbol{\Phi_j}^{0},\boldsymbol{Z}^{0},\boldsymbol{U}^{0},{\omega_{1,m}^{(t)}}$, ${\omega_{2}^{(t)}}$;
            \While
                \State\  $\boldsymbol{\Phi}_j^{k+1}=\underset{\boldsymbol{\Phi_j}}{\mathop{\arg\min}}\ \mathcal{L}_{\rho}(\boldsymbol{\Phi}_j,\boldsymbol{Z}^k,\boldsymbol{U}^k)$ with proximal gradient approach
                \State $\boldsymbol{Z}^{k+1}$ update via equation (\ref{equ:Z update})
                \State $\boldsymbol{U}^{k+1}$ update via equation (\ref{equ:U update})
                \If{$\big|\mathcal{L}_{\rho}(\boldsymbol{\Phi_j}^{k+1},\boldsymbol{Z}^{k+1},\boldsymbol{U}^{k+1})-\mathcal{L}_{\rho}(\boldsymbol{\Phi_j}^{k},\boldsymbol{Z}^{k},\boldsymbol{U}^{k})\big|\leq \epsilon$}
                    \State break;
                \EndIf                
                \State $k = k+1$
            \EndWhile
            
            \Return $\boldsymbol{\hat{\theta}_{j}}^{(1)},...,\boldsymbol{\hat{\theta}_j}^{(T)}$
        \EndFor
    \end{algorithmic}
\end{algorithm}

\subsection{Network construction}

With the modeling, estimating, and computing methods developed above, we can construct and visualize the proposed dynamic HDN on costs. Then key properties such as adjacency, connectivity, and module can be
summarized and analyzed.\\

\textbf{Adjacency}: The network structure is captured by a $d \times d$ adjacency matrix where a non-zero element indicates conditional dependency of corresponding nodes. Under the dynamic framework, we construct one adjacency matrix for each time period based on estimated $\beta_{i,j}^{(t)}$ and $\eta_{i,j}^{(t)}$. Specifically, the adjacency matrix  $\boldsymbol{A}^{(t)} = [a_{ij}^{(t)}]$ is defined as

\begin{equation}
    \label{equ:adjacency}
    \begin{aligned}
        a_{ij}^{(t)}=\begin{cases}
                0,\quad\ {\rm if}\ \beta_{i,j}^{(t)}<\kappa\ \&\ \eta_{i,j}^{(t)}<\kappa\ \&\ \beta_{j,i}^{(t)}<\kappa\ \&\ \eta_{j,i}^{(t)}<\kappa,\\
                1,\quad\ {\rm otherwise},\\
                \end{cases}
    \end{aligned}
\end{equation}

where the threshold $\kappa$ is utilized to further remove trivial connections, thereby producing a network that is sparse and interpretable. The choice of $\kappa$ can be data driven or based on literature. In our analysis, $\kappa$ is chosen to achieve an average node connectivity comparable to the average number of CCS diseases per person in real data.\\

\textbf{Connectivity}: Defined as the number of other nodes connected to a specific node, connectivity measures that node's influence within the network.
For a given node $j$, its connectivity at time $t$ is defined as
\begin{equation}
    \label{equ:connectivity}
    C_j^{(t)}=\sum_{i\neq j}a_{ij}^{(t)}.
\end{equation}

Nodes with higher connectivity tend to play a more central role in the network, a point emphasized in various studies on network analysis. \citep{mei2023human}\\

\textbf{Module}: Also known as network clusters or communities, modules are subsets of the network where nodes are densely interconnected. The topological overlap matrix (TOM) is adopted for module construction. The $ij$th element of the TOM at time $t$ is defined as 

\begin{equation}
    \label{equ:module}
    \text{TOM}_{ij}^{(t)} = \frac{e_{ij}^{(t)}+a_{ij}^{(t)}}{min\{C_i^{(t)},C_j^{(t)}\}+1-a_{ij}^{(t)}},
\end{equation} 

where $e_{ij}^{(t)} = \sum_{u=1}^{d} a_{iu}^{(t)} a_{uj}^{(t)}$ counts the shared neighbors between nodes $i$ and $j$. The dissimilarity between nodes is then expressed as $\text{dissTOM} = 1 - \text{TOM}$. Using this dissimilarity metric, we apply hierarchical clustering combined with dynamic tree cutting to identify distinct modules within the network. \citep{yip2006generalized}\\

\section{Simulation}

A simulation study is conducted to evaluate the performance of the proposed model. We explore a range of scenarios with diverse graph structures (spanning from entirely random graphs to fixed graphs), network sparsity levels, and time-dependent mechanisms. We choose the number of nodes to be \( d \in \{100, 125, 150\} \) and a sample size of \( n = 10,000 \) to closely represent the real dataset. 
Given the requirement for smoothness, a longer time period is generally preferable. We conservatively adopt a time period count of \( T = 5 \).
This results in a total of $36$ simulation settings. Details can be found in Table. \ref{tab:sim F1} Simulated values of $Y_{j}^{(t)}$ and $V_{j}^{(t)}$ ($j=1,...,d;t=1,...T$) are generated based on distributions defined through Equation (\ref{equ:distribution}) with Gibbs sampling. Relevant non-zero parameters are randomly chosen from pre-specified distributions detailed in Table A.1 of the Supplementary Materials. \\

Network structure at each time period is fully determined by the parameters $\boldsymbol{\beta}_{j}^{(t)}$ and $\boldsymbol{\eta}_{j}^{(t)}$. Define two $d\times d$ matrices $M(\boldsymbol{\beta}^{(t)})$ and $M(\boldsymbol{\eta}^{(t)})$ with non-zero parameters in the matrices indicate the presence of connections between specific nodes, we consider three initial graph structures: band structure, Watts Strogatz (WS) network, and stochastic block model (SBM). The band structure is a fixed type of graph structure in which the connections between nodes are limited to a specified bandwidth. SBM represents a random graph where the vertex set can be divided into $K$ groups, typically used to represent different communities, with nodes between each community connected with different probabilities. \citep{holland1983stochastic} WS is a graph structure between completely regular and random networks, characterized by high clustering coefficients and short average path lengths. \citep{watts1998collective} For all three types of graph structures, sparsity level can be adjusted by tuning relevant parameters. The sparsity of matrices $M(\boldsymbol{\beta}^{(t)})$ and $M(\boldsymbol{\eta}^{(t)})$ reflects the connectivity of nodes in the proposed network. Based on existing HDN literature, we consider two sparsity levels $5\%$ and $15\%$. As the diagonal elements in the matrices $M(\boldsymbol{\beta}^{(t)})$ and $M(\boldsymbol{\eta}^{(t)})$ have no practical meaning, they are set to $1$ for convenience.\\
 
To generate time-dependent datasets after setting the initial graph structure, we consider two different types of time evolution mechanisms, one with small and continuous changes and one with a one time sudden change. The first mechanism involves randomly selecting edges to add or remove, causing the graph to continuously change over time. Specifically, the number of changing edges is set to 12\% of the current edge count per year, with a 50\% probability of adding a new edge and a 50\% probability of removing an existing edge. The second mechanism involves randomly selecting a subset of nodes at a specific time period and reversing all their connections. Specifically, 5\% of the nodes are reversed at $t=3$ to induce discontinuous changes. \\

To better evaluate the performance of the proposed approach, we compare it with two alternative methods. The first alternative constructs static networks that account for zero-inflation, as proposed by Mei et al. (2023), \cite{mei2023human} on a year-by-year basis. The second alternative employs a Dynamic Bayesian Network. \citep{song2009time} Other time-dependent modeling approaches exist for network analysis, including the well-known Time-Varying Graphical Lasso \citep{hallac2017network} and Multivariate Autoregressive Models. \citep{hytti2006tutorial} Preliminary results (omitted here) show that these methods perform similarly in our setting. We select the Dynamic Bayesian Network as a representative example of time-dependent methods that cannot accommodate zero-inflation. A direct comparison of our method with these two alternatives effectively demonstrates the advantages of incorporating both the dynamic feature and the zero-inflation nature of the data. \\

\subsection{Simulation results}
The precision of different methods is evaluated using the F-score for identifying correct edges. F-scores are calculated for each time period and then averaged across the five periods. As shown in Table \ref{tab:sim F1}, the proposed method outperforms the two alternatives in all scenarios, providing more accurate estimates and greater stability. 
It is also observed that the F-scores of the proposed method do not decrease as $d$ increases. This is because a sufficiently large sample size has been selected, even at the highest level of $d$. We would expect the F-scores to decline if we keep increase $d$ with the sample size fixed, but this aspect is beyond the scope of our simulation. The chosen values for $d$ and $n$ are intended to closely mirror the dimensionality of the real-world data, and the simulation results demonstrate that the sample size we use to work with the real dataset is adequate.\\

While the traditional time-dependent method fails to identify correct edges due to zero-inflation, the static network constructed on a year-by-year basis performs only slightly worse than our method in edge detection. However, temporal smoothing, as incorporated in our method through the adaptive fused LASSO penalty, reduces noise and enhances the detection of underlying trends, leading to more reliable analysis and improved predictive accuracy. To evaluate the impact of the adaptive fused LASSO penalty, we define

\begin{equation}
    \Delta=\frac{1}{T-1}\sum_{t=2}^T\sum_{j=1}^d||\boldsymbol{\hat{\theta}}_{j}^{(t)}-  \boldsymbol{\hat{\theta}}_{j}^{(t-1)}||_1,
\end{equation}

to represents the average difference between estimated parameters in consecutive years. This metric is used to compare our method with the static network approach. Given that under the time-varying mechanism 2, the network undergoes a sudden change in the third year, we further divide $\Delta$ into two parts:

\begin{equation}
    \begin{aligned}
        &\Delta_1 = \sum_{j=1}^d||\boldsymbol{\hat{\theta}}_{j}^{(3)}-  \boldsymbol{\hat{\theta}}_{j}^{(2)}||_1,\\
        &\Delta_2 = \frac{1}{T-2}[(T-1)\Delta - \Delta_1].\\
    \end{aligned}
\end{equation}
$\Delta_1$ represents the average estimated parameter changes from year 2 to year 3, while $\Delta_2$ denotes the average estimated parameter changes for the remaining years. As shown in Table \ref{tab:sim delta}, under time-varying mechanism 1, where the network experiences small and continuous changes, our method achieves smooth and stable estimates with significantly smaller values of $\delta$ and lower standard deviation. Under time-varying mechanism 2 where the network undergoes a sudden change in year 3, our method results in comparatively large values of $\Delta_1$ and smaller values of $\Delta_2$. This indicates that the proposed temporal smooth approach can accurately capture both small, continuous variations and large, sudden changes.
\\


\section{Data Analysis}
The proposed model is applied to Taiwan's claims data, which are sourced from a single-payer, mandatory enrollment insurance program that covers 99.9\% of the Taiwanese population. It is estimated that almost all hospital and clinic-based health care in Taiwan is facilitated through the program. \citep{hsieh2019taiwan}
This database is considered uniquely beneficial due to its comprehensive coverage and superior quality. It is the most authoritative and representative database of the unique Taiwanese population.

\subsection{Data preparation}

The analyzed dataset consists of 5,849,402 claims, including both inpatient and outpatient claims. These records are derived from a randomly selected sample of 10,000 eligible subjects from the Taiwanese population for the period of 2000 to 2013. 
The sample size necessary for accurate estimation depends on the complexity of the network.
In our simulation, the proposed method demonstrates stable and reliable results with a sample size of 10,000 for 125 nodes across various network structures.
Therefore, for real data analysis, we consider 10,000 as a sufficient sample size to ensure both computational feasibility and reliable results.\\

\begin{figure}[htbp]
    \centering
    \captionsetup{justification=centering}    
    \includegraphics[width=0.8\linewidth]{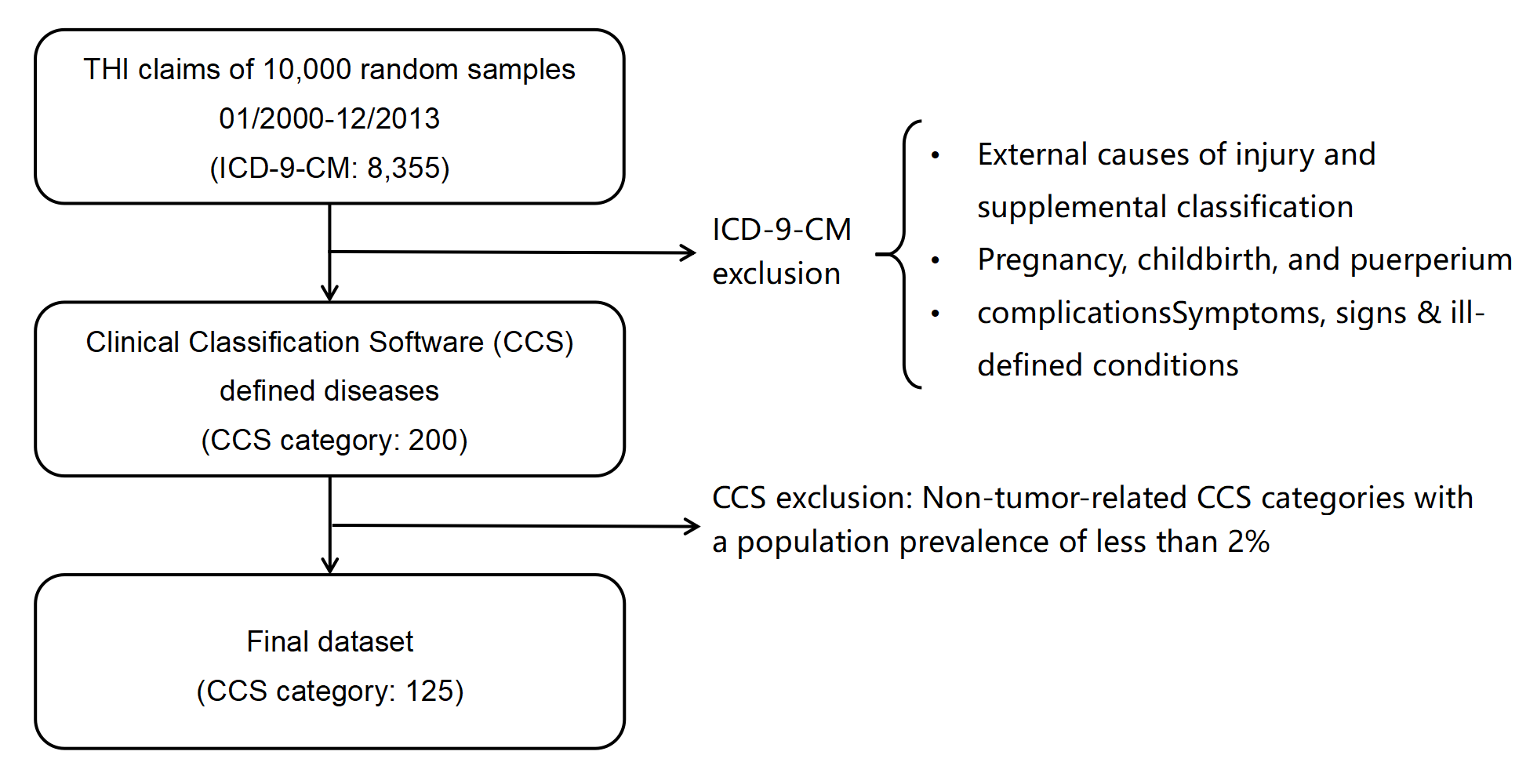}
    \caption{Flowchart of data processing}
    \label{fig:flowchart}
\end{figure}

Diseases are defined by diagnosis codes under the International Classification of Diseases, Ninth Revision, Clinical Modification(ICD-9-CM). The ICD-9-CM code version 1992 (used before 2005) is transformed into the 2001 version. Following the literature, \citep{ma2020human} we exclude the following ICD-9-CM codes from analysis: external causes of injury and supplemental classification (the E and V codes), pregnancy, childbirth and puerperium complications (630 – 679), and symptoms, signs \& ill-defined conditions (760-999).
This leads to 7333 ICD-9-CM codes.
In order to better define human diseases, these diagnosis codes are further grouped into more clinically meaningful categories using the Clinical Classifications Software(CCS), developed by the U.S. Agency for Healthcare Research and Quality. \citep{mei2023human} 
To generate accurate estimates, we exclude non-tumor-related rare diseases with a population prevalence of less than 2\% over the study period. This leads to 125 CCS categories for final analysis.
The flow of diagnosis processing is sketched in Figure. \ref{fig:flowchart}\\

In our analysis, we focus on the diseases treatment cost, which involves summing the costs for each person and each disease across all inpatient and outpatient treatment episodes within a specific time period.
It is important to note that a single hospital visit may involve multiple diseases.
For each claim, there can be up to five diagnosis codes (one primary and four secondary) diagnosis codes. 
We attribute 60\% of the costs associated with each claim to the primary diagnosis and evenly distribute the remaining 40\% among all recorded secondary diagnoses.
This adopted allocation method ensures that the primary diagnosis is prioritized during each visit, while all coexisting conditions are treated with equal importance, as their order of occurrence often lacks specific clinical significance. 
It is also worth mentioning that there is no consensus in the existing literature on the optimal allocation method for multiple disease conditions. Sensitivity analyses are conducted by allocating 55\% and 70\% of per-claim costs to the primary diagnosis, and the results are consistent (detailed results omitted). 
To eliminate the impact of currency inflation, medical costs are adjusted using the Consumer Price Index of Taiwan (data derived from the \href{www.imf.org/en/Publications/WEO}{World Economic Outlook database}). This adjustment accounts for changes in the purchasing power of money over time by converting historical costs to their 2013 equivalent values.\\


In addition to Figure \ref{fig:discriptive}, we present further descriptive results in the Supplementary Materials. Figure A.1 shows the histograms of the number of claims and the number of CCS diseases per person. On average, one patient has 471 claims and 17 CCS diseases. Figure A.2 displays the top 10 diseases with the highest prevalence and per capita medical costs. Prevalence and costs illustrate the disease-specific medical burden from different perspectives. 
It is observed that chronic conditions generally exhibit higher prevalence, whereas cancers predominantly feature among the top 10 in terms of costs. Additionally, diseases with high prevalence show less temporal variation across the study window compared to diseases with high costs.
\\

\subsection{Network analysis}

Analyzing Taiwan's claims data from 2000 to 2013 using the proposed model resulted in a series of time-dependent yearly networks. Each network is graphically visualized using the software \textit{Gephi} and presented in Figure A.3 in the Supplementary Materials. In these plots, the size of a node is proportional to its connectivity, and the color of a node indicates its module membership.
Table \ref{tab:edges and modules} presents the number of edges and the number of modules for each yearly network constructed using our method, as well as the two alternative methods described in the Simulation section. 
Since the Dynamic Bayesian Network uses the outcomes from the previous year to predict the outcomes for the following year, the results for Year 2000 under Alternative 2 are not applicable.
All methods exhibit noticeable temporal variations; however, the results differ significantly. Our method yields fewer edges compared to the static network approach, but more edges than the Dynamic Bayesian Network approach. 
The number of modules is similar across the three methods, but our method produces smoother and more stable results over time.
Given the large number of diseases, interconnections, and temporal variations, it is challenging to discuss all findings in detail. In what follows, we will focus on key findings, representative examples, and their potential implications for healthcare settings.\\

\subsubsection{Connectivity}

\begin{figure}[htbp]
    \centering
    \captionsetup{justification=centering}
    \includegraphics[width=0.85\linewidth]{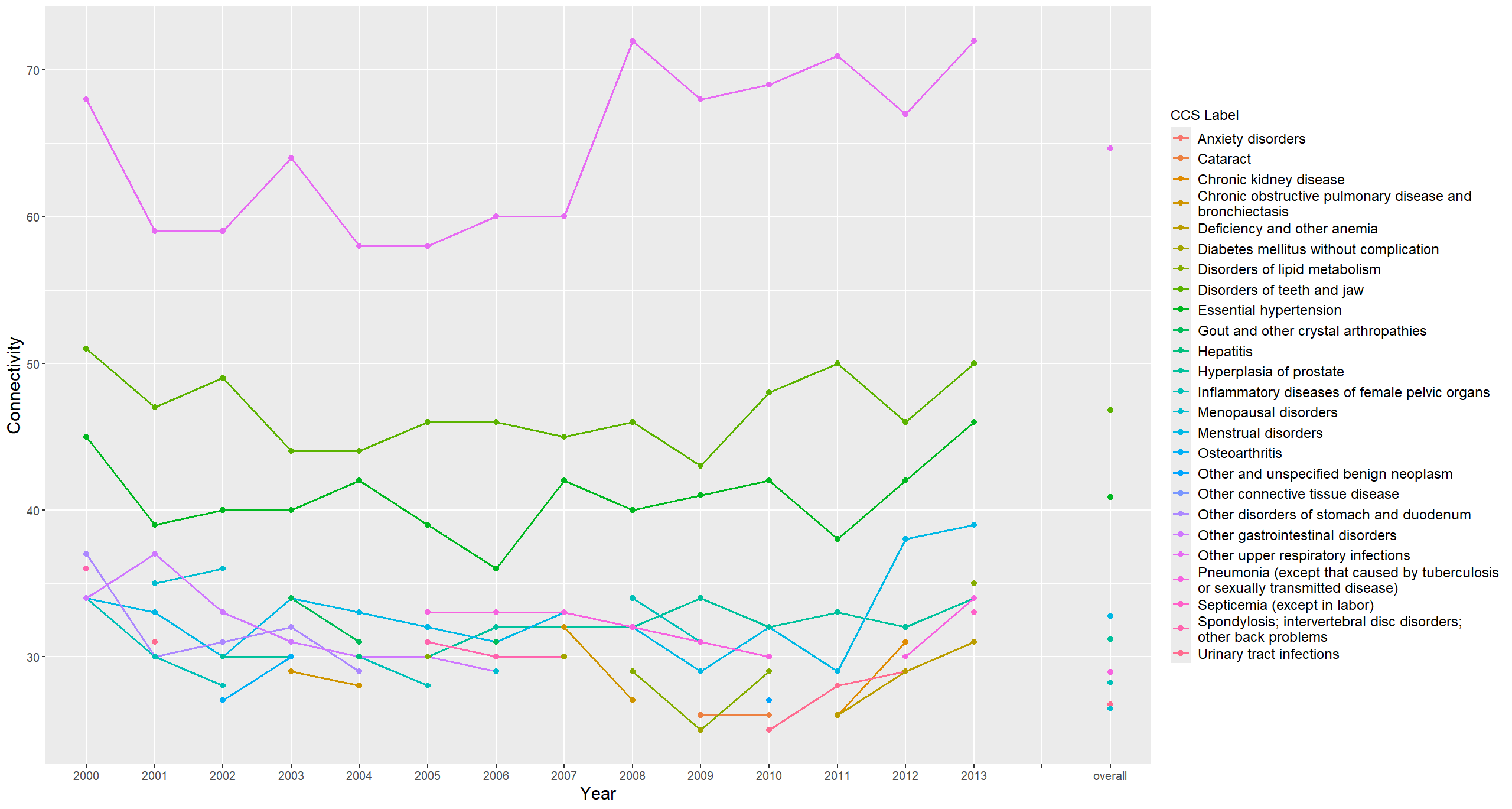}
    \caption{Top ten diseases with the highest connectivity}
    \label{fig:10conn}
\end{figure}

Figure \ref{fig:10conn} displays the top 10 diseases with the highest connectivity, both annually and overall. These diseases are prevalent and well-studied, known to be associated with significant medical burdens. A comparison of Figure \ref{fig:10conn} with Figure A.2 reveals that diseases with higher prevalence rates generally exhibit greater connectivity and, consequently, a larger impact within the network. This observation may offer valuable insights for future policy development.
As depicted in the graph, yearly disease connectivity experiences fluctuations, without a discernible trend over time. Notably, diseases that are ranked top for yearly connectivity exhibit a high degree of overlap across years, with their ranking remaining relatively stable. This suggests that the temporal variations in connectivity are insubstantial. \\

\subsubsection{Module}

\begin{figure}[htbp]
    \centering
    \begin{minipage}[c]{0.6\linewidth} 
        \centering
        \subfloat[Time-varying HDN on medical costs: Year 2010]{
            \includegraphics[width=\linewidth]{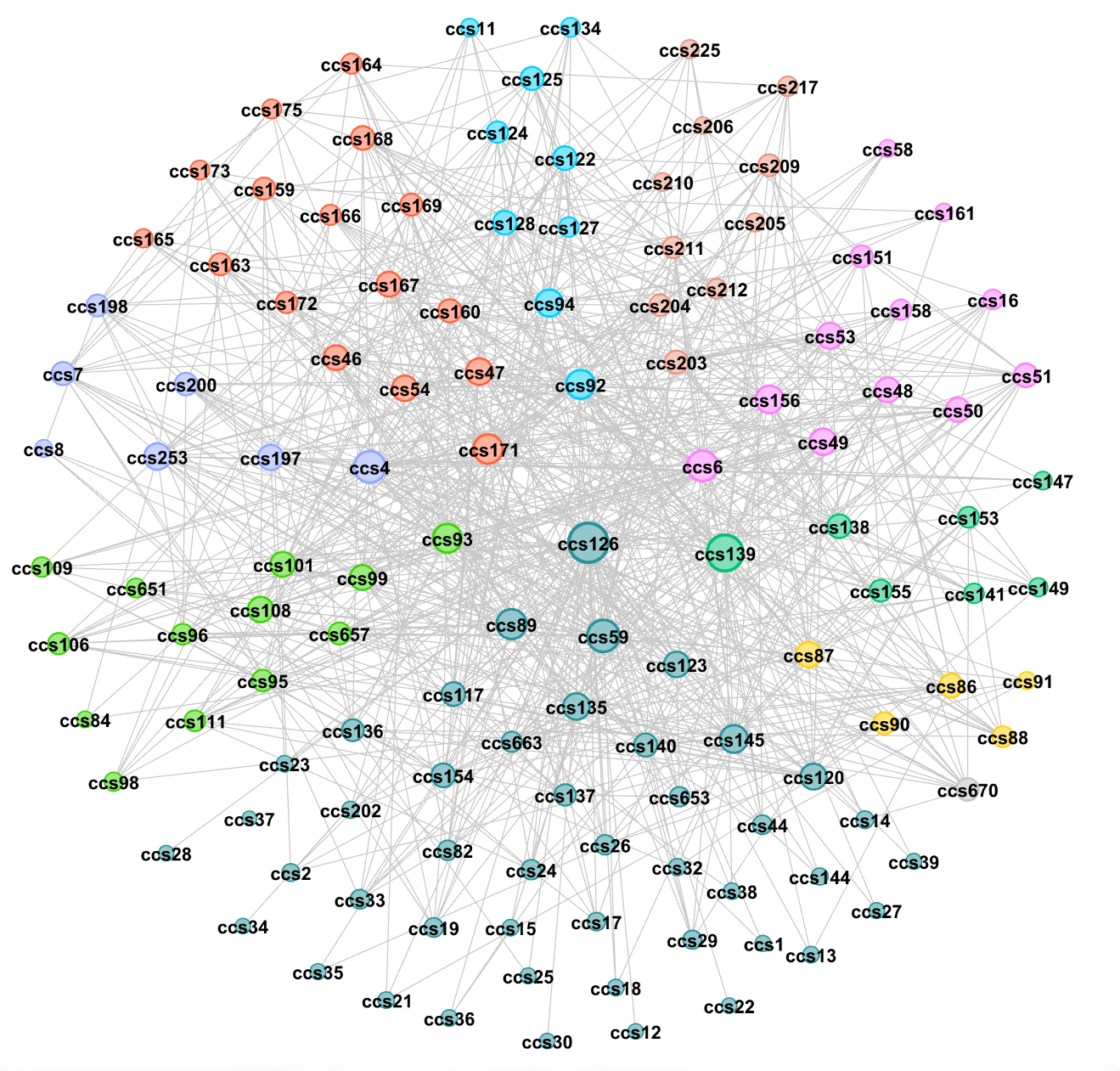}
            \label{fig:module_2010}
        }
    \end{minipage}%
    \begin{minipage}[c]{0.4\linewidth} 
        \centering
        \subfloat[Hub diseases: Year 2010]{
        \includegraphics[width=0.95\linewidth]{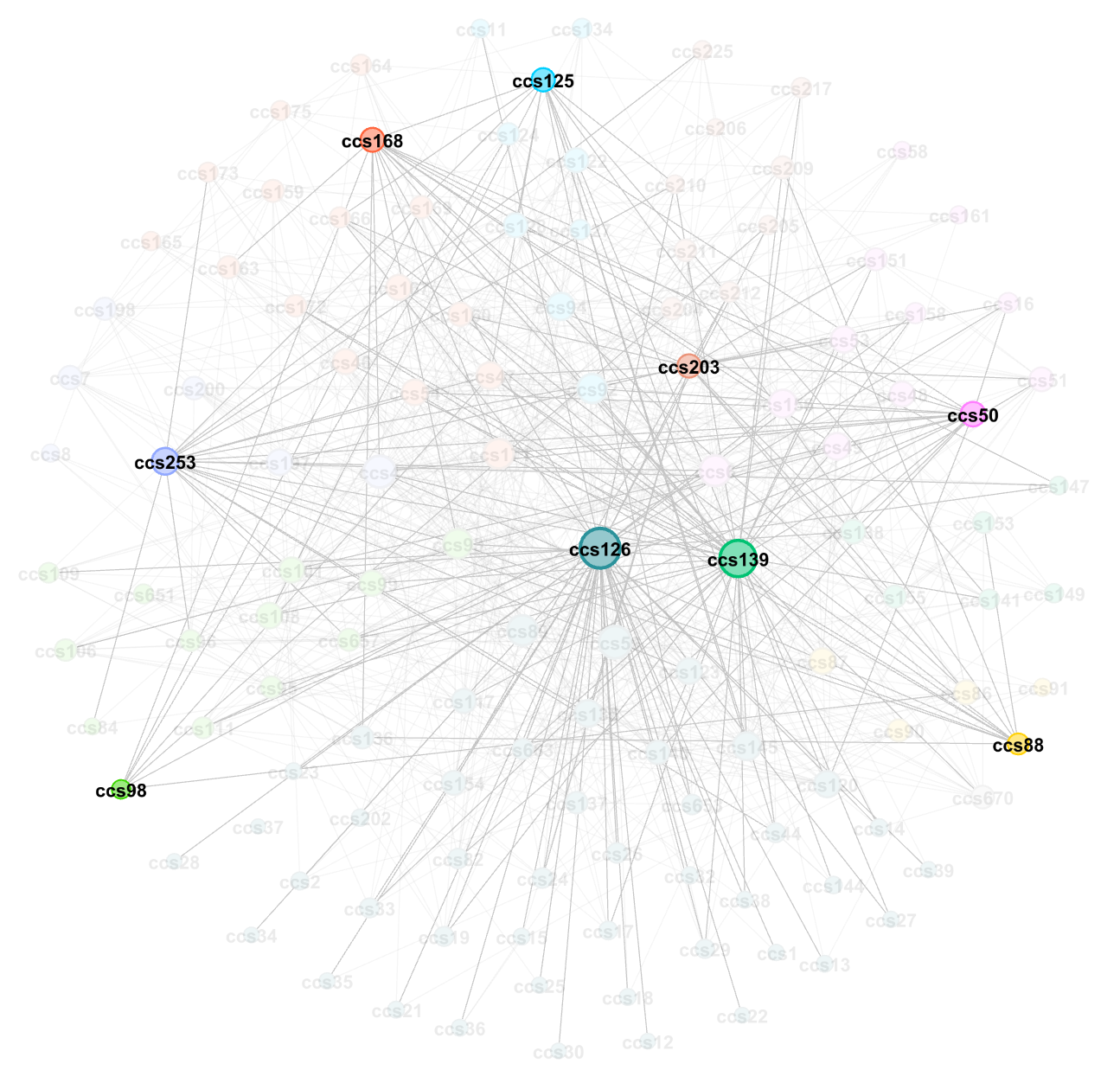}
            \label{fig:hub_2010}
        } \\
        \subfloat[Yearly change in pairwise disease module relationships]{
            \includegraphics[height=1.3in, width=\linewidth]{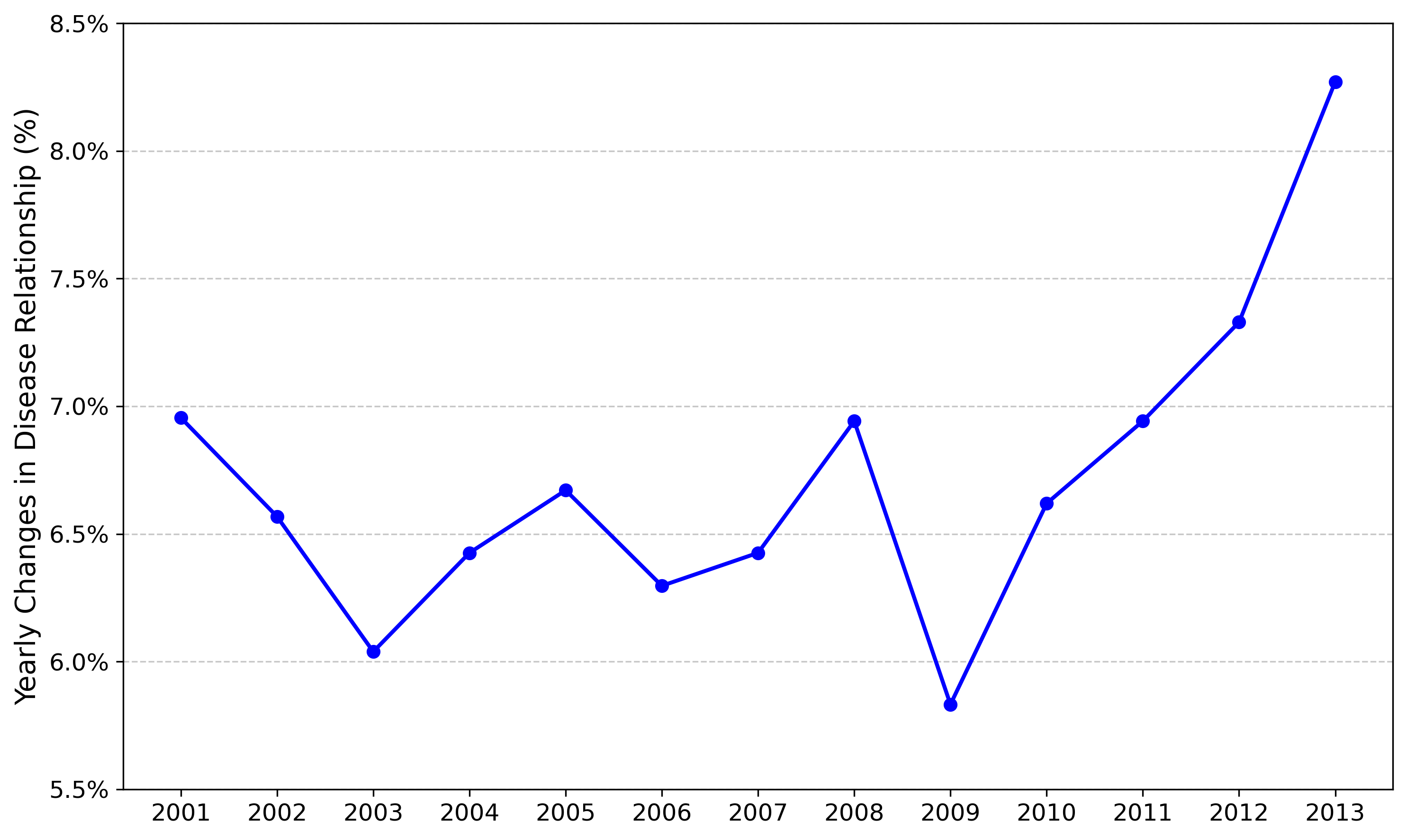}
            \label{fig:module_chg}
        }
    \end{minipage}
    \caption{Disease modules and their temporal variations}
    \label{fig:module}
\end{figure}

For module interpretation, Figure \ref{fig:module_2010} shows an representative example of the network of Year 2010, which comprises 125 nodes and 884 edges, with these nodes being categorized into 9 modules containing between 5 to 45 diseases each. Upon detailed examination of these modules, they are found to correspond to specific disease categories as follows: cancers and complications (navy), reproductive and urogenital diseases (orange), cardiovascular and cerebrovascular diseases (light green), endocrine and metabolic diseases (pink), bone and connective tissue diseases (brown), respiratory diseases (blue), dermatological diseases (purple), gastrointestinal diseases (green), ocular diseases (yellow). Extensive studies have explored the interconnections of diseases within each module from the etiology and pathology perspectives, further suggesting the significance of the proposed network. While these studies focus on a limited number of pre-selected diseases, our module analysis offers pan-disease level evidence of disease classification. Analyzing the constructed modules, we can more accurately identify factors influencing a group of closely interconnected diseases, thereby supporting multimorbidity management strategies.\\

Connectivity and intramodular connectivity describe the importance of diseases from different perspectives. For example, benign prostatic hyperplasia (BPH) is central to the urogenital disease module with the second highest intramodular connectivity, but has relatively low overall connectivity. This is consistent with existing etiology research findings, as disorders of the urogenital system are often relatively independent and usually not significantly related to disorders of other systems. \citep{BouZerdan2022} Through the network analysis in this study, these diseases have been effectively grouped, with BPH being identified to have the greatest local impact within the module. This finding can inform interventions targeting urogenital diseases. According to the literature, although age and sex hormones are considered the two major risk factors for BPH, its exact etiology remains unclear. \citep{langan2019benign} Therefore, in-depth research on diseases closely related to BPH may provide more insights into this condition.\\

Temporal variation in modules has been assessed by alterations in the pairwise module relationships. Specifically, when two diseases are classified within the same module one year but fall into different modules the following year, we consider their module relationship to have changed. This applies in reverse as well. Figure \ref{fig:module_chg} illustrates the annual percentage of disease pairs with changed module relationships. From 2001 to 2011, there are minor changes (5.5\% - 7\%), indicating a relatively stable disease module membership. Notable increases in changes were observed in 2012 and 2013, likely due to the upgrade of program infrastructure and the reform of health policies to expand coverage and benefits. \citep{lai2018critical}\\

\subsubsection{Temporal variation}

\begin{figure}[htbp]
    \centering
    \captionsetup{justification=centering}
    \subfloat[Edge changes from 2000 to 2001]{
    	\includegraphics[width=0.4\textwidth]{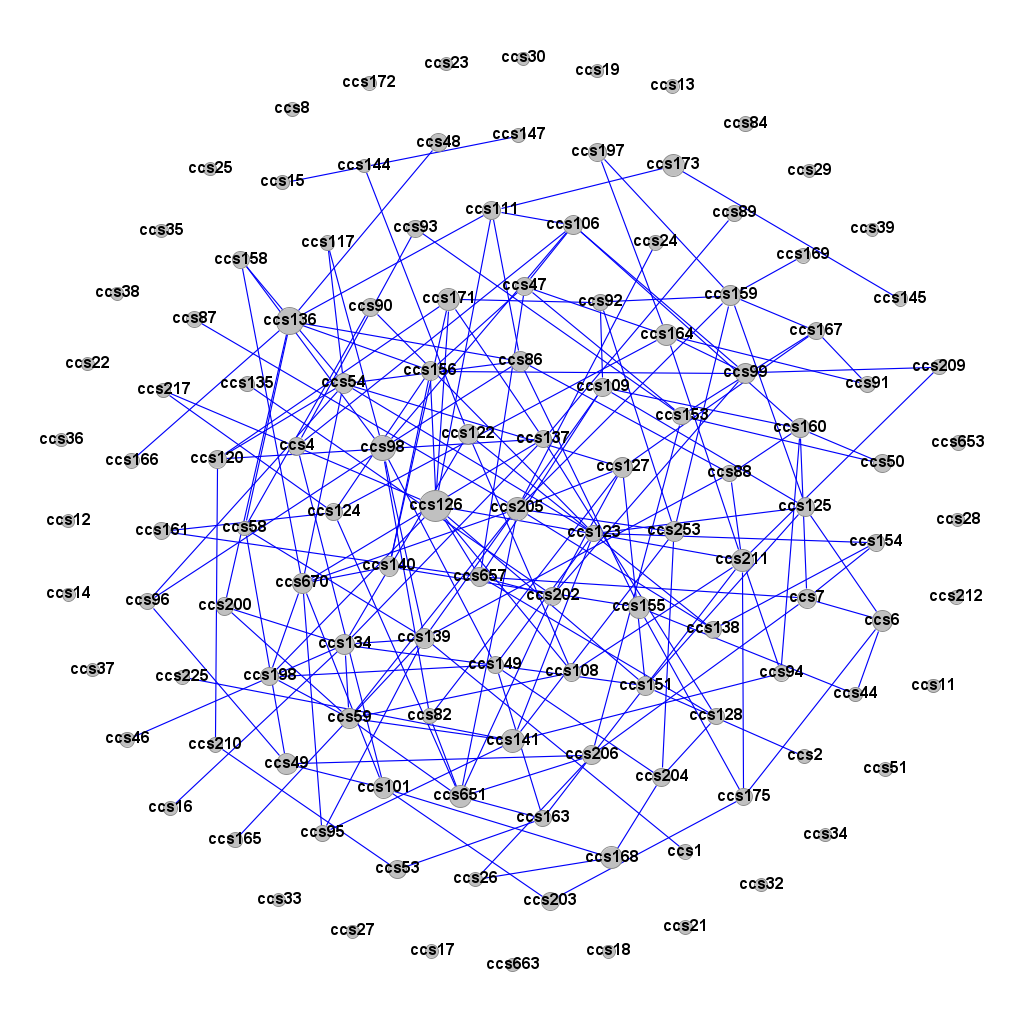} 
    	\includegraphics[width=0.4\textwidth]{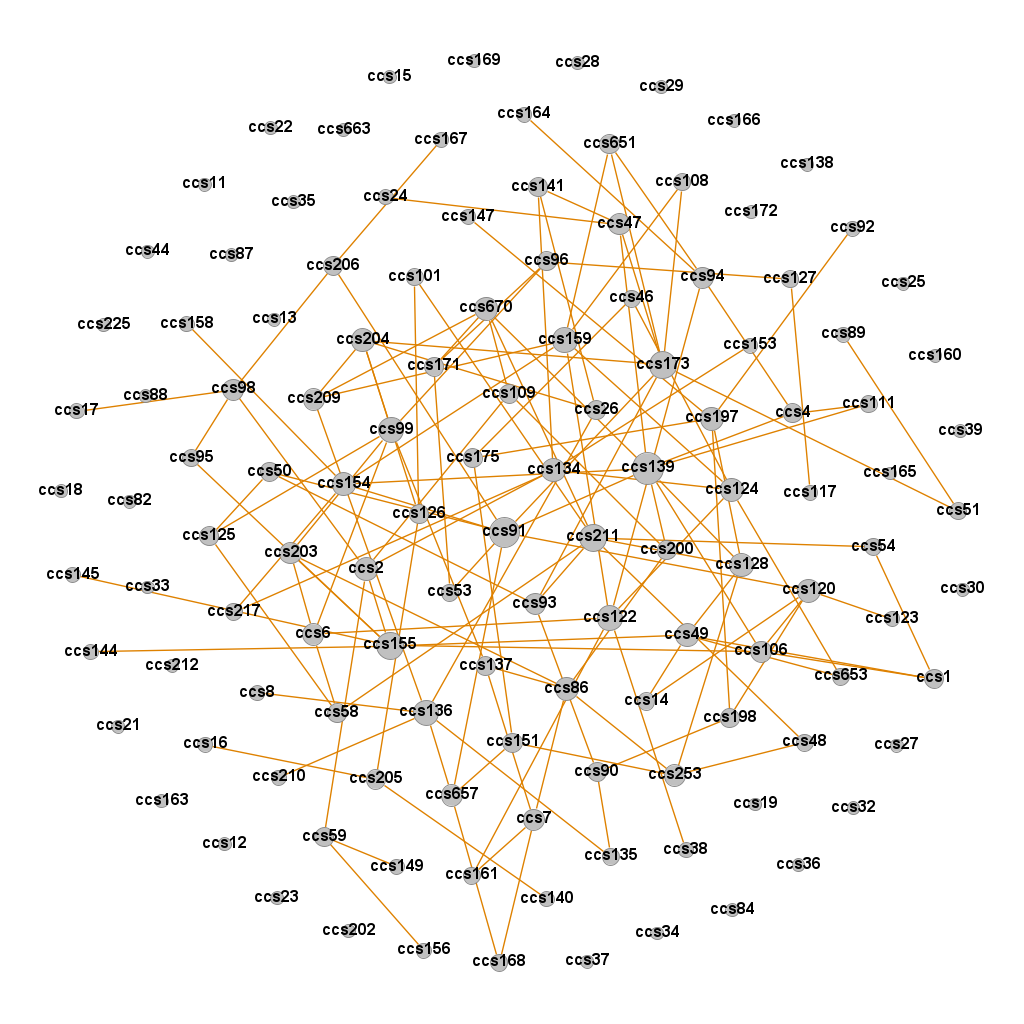} 
     \label{fig:edge_chg1}
     }
     
     \subfloat[Edge changes from 2012 to 2013]{
    	\includegraphics[width=0.4\textwidth]{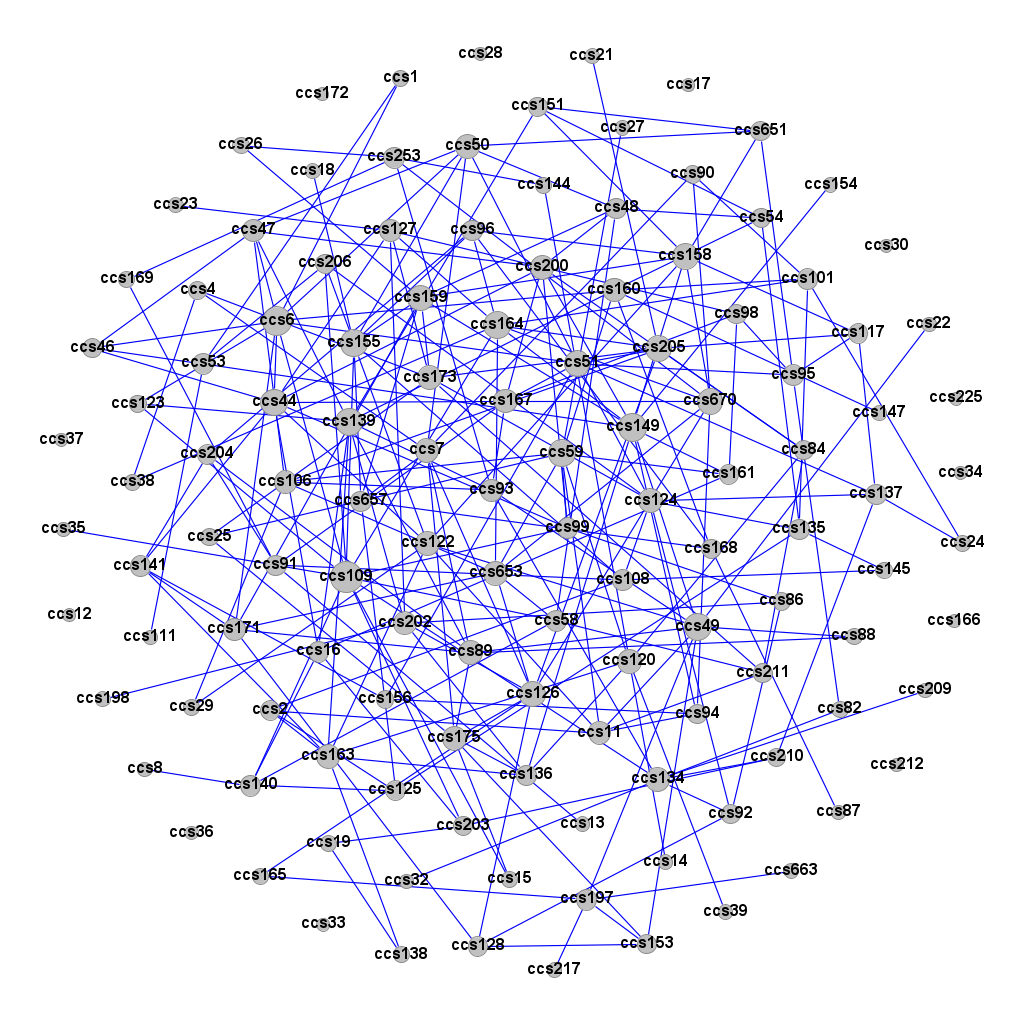} 
    	\includegraphics[width=0.4\textwidth]{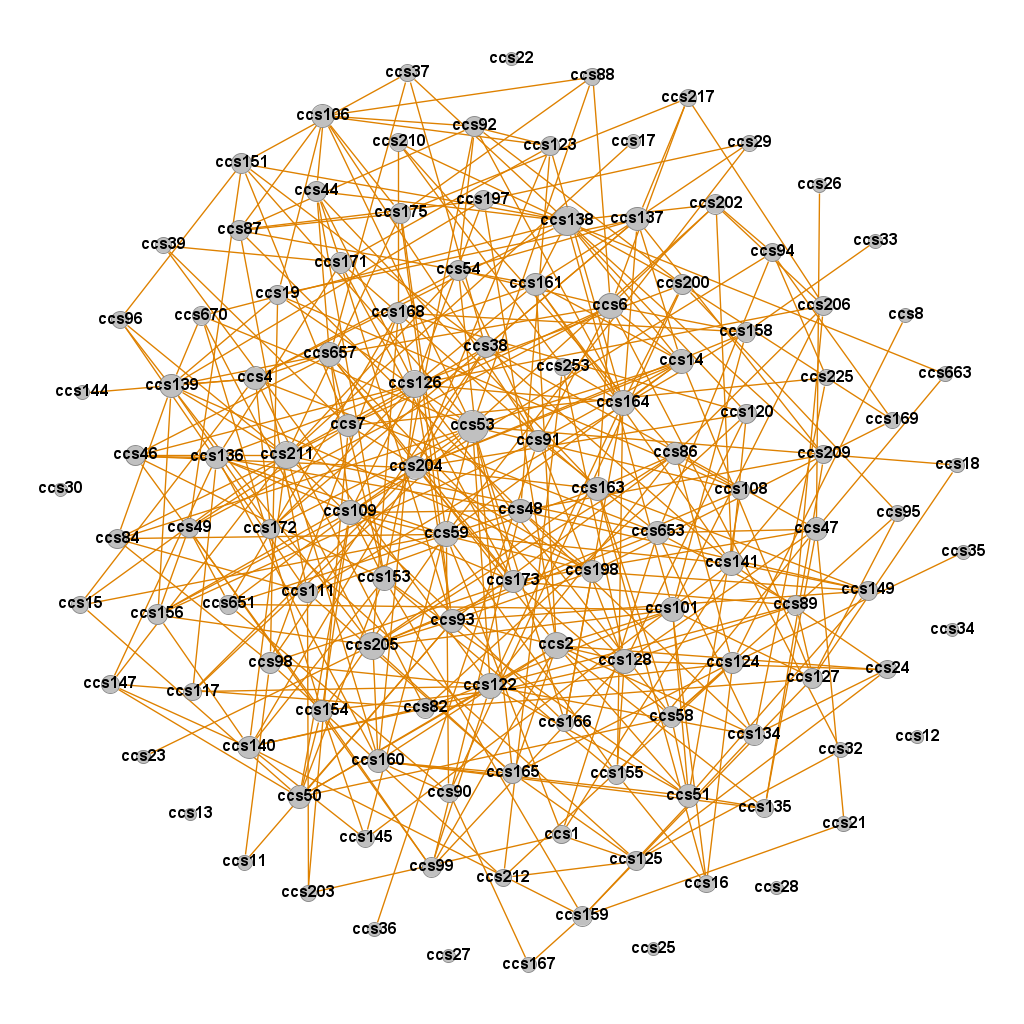}
     \label{fig:edge_chg2}
     }
     
    \caption{Edge changes between two consecutive years in different periods}
    \label{fig:edge_chg}
\end{figure}

As illustrated in simulation, the adoption of the adaptive fused LASSO penalty facilitates the detection of both minor, continuous variations and major, sudden changes. Figure \ref{fig:edge_chg} presents edge changes between two consecutive years in different periods. Edges that are present in the earlier year but absent in the later year are highlighted in blue, while the reverse situation is highlighted in orange. Unchanged edges are shown in grey. Figure \ref{fig:edge_chg1} illustrates the edge changes from 2000 to 2001, representing a case of minor changes. In 2000, 16.83\% of edges disappeared by 2001, while 15.79\% of edges in 2001 were new additions not present in the previous year. Figure \ref{fig:edge_chg2} shows the edge changes from 2012 to 2013, representing a case of significant changes. In 2012, 26.46\% of edges disappeared by 2013, while 35.41\% of edges in 2013 were new additions not present in the previous year. The substantial changes observed from 2012 to 2013 are also likely attributable to the earlier-mentioned reforms in Taiwan's health insurance program.

\subsubsection{An example disease - breast cancer}

To better illustrate how the proposed network can assist multimorbidity management, we further discuss the network features of breast cancer (CCS24) as a representative example. Breast cancer, characterized by its high prevalence and significant costs, often necessitates comprehensive management of common comorbidities and complications arising from treatments. \citep{runowicz2016american}\\

Table \ref{tab: breast cancer} lists diseases that are frequently ($\geq 5$ years) connected to breast cancer and are consistently ($\geq 5$ years) clustered with breast cancer in the same module across years. These diseases should be be taken into serious consideration when designing treatment, health management, and insurance strategies for individuals with breast cancer. Many of the diseases listed in Table \ref{tab: breast cancer} have been previously identified as closely related to breast cancer, highlighting the significance of our findings.
For example, non-malignant breast conditions and breast cancer are interconnected every year over the span of 14 years. This is expected, as non-malignant breast conditions are the most common findings during breast cancer screenings. \citep{AndersJohnson2009} Multiple malignant cancers (eg., Hodgkin's disease, Cancer of thyroid, and Leukemias) are frequently clustered with breast cancer and studies have shown that they share the same etiological, pathological, and clinical risk factors. \citep{ferrari2014expression,joseph2015association,vaziri2020leukemia} 
It is noted that conditions that are frequently connected to and clustered with breast cancer are not entirely consistent, indicating the importance of studying both direct and indirect interconnections.\\

The presence of certain ``unexpected'' diseases provides additional insights. For instance, periodontal disease is interconnected with breast cancer across all 14 years. Although the specific mechanisms of this association remain unclear, studies have shown that periodontal disease is linked with various tumors, including gastrointestinal malignancies, lung cancer, and breast cancer. \citep{fitzpatrick2010association} Further examination of the network structure involving periodontal disease may offer valuable information for future etiological and pathological studies. Additionally, it is noteworthy that breast cancer clusters with cancers that predominantly occur in males, such as testicular cancer and prostate cancer. While breast cancer can occur in men, it is exceedingly rare. Research has identified a familial pattern in the occurrence of breast and other genital cancers, which may be attributable to shared genetic risk factors inherited within families. \citep{risbridger2010breast} In the constructed network, genital cancers, including breast cancer, gynecological cancers, testicular cancer, and prostate cancer, frequently cluster within the same module. These conditions are all hormone-dependent, and our analysis could potentially lead to common treatment and health management strategies for this group of hormone-related cancers.\\

\begin{figure}[htbp]
    \centering
    \captionsetup{justification=centering}
    \includegraphics[width=0.45\linewidth] {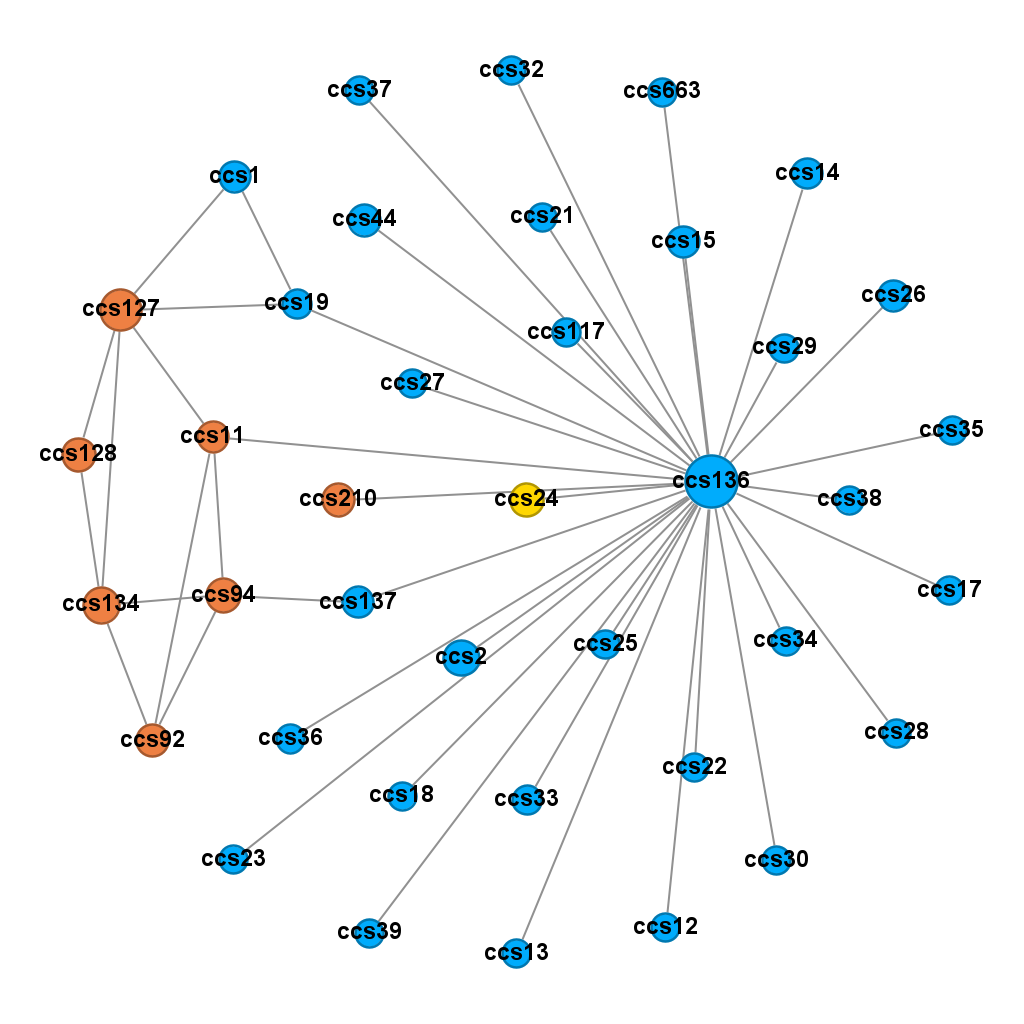}
        \hspace{0.75cm}
	\includegraphics[width=0.45\textwidth]{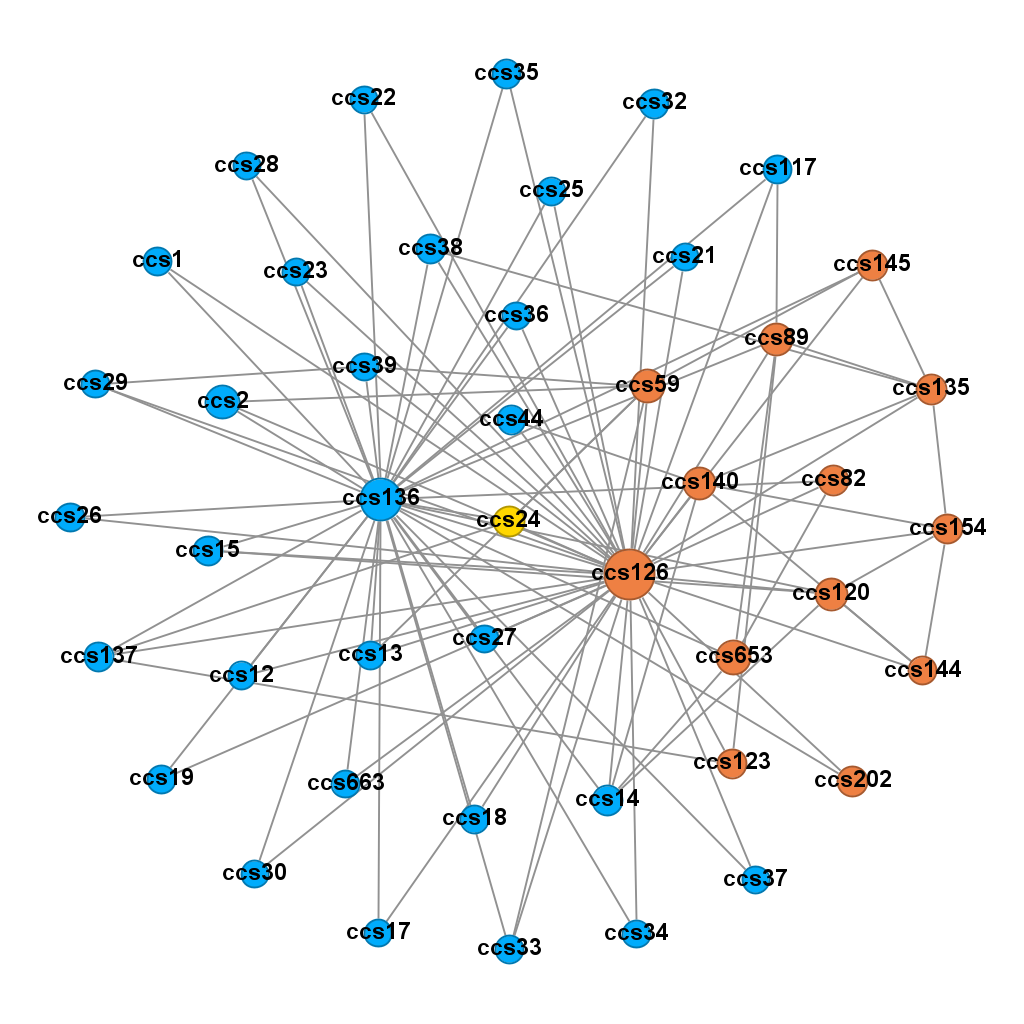}
    \caption{Modules that contain breast cancer in 2003 (left) and 2010 (right)}
    \label{fig:ccs24_module_chg}
\end{figure}

A further examination of the temporal variation in the network structure of breast cancer reveals clinically relevant insights. Figure \ref{fig:ccs24_module_chg} illustrates the modules to which breast cancer (highlighted in yellow) belongs in 2003 and 2010. Nodes colored in blue are shared between the two yearly modules, while nodes in orange appear in only one of the two years.
In the 2000s, significant advances were made in breast cancer treatment. Before 2003, chemotherapy was the primary treatment for breast cancer, while numerous targeted drug therapies emerged thereafter. \citep{meric2019advances} Consequently, substantial changes in the network structure of breast cancer are anticipated. For example, gastrointestinal reactions are common adverse effects associated with small molecule anti-HER2 agents. \citep{mukai2010targeted} Thus, in the 2010 network, we observed that breast cancer and several gastrointestinal diseases (intestinal infection, gastritis and duodenitis, and regional enteritis and ulcerative colitis) are clustered together, whereas these conditions are absent in the 2003 module.
Certain diseases that are in the same module as breast cancer in 2003 are no longer clustered with breast cancer in 2010, possibly due to changes in treatment strategies. For instance, systemic lupus erythematosus (SLE) and breast cancer, which predominantly affect females, have seen efforts focused on treatment strategies to minimize adverse effects when both conditions are present. \citep{kontos2008systemic} The change in module membership for SLE and breast cancer between 2003 and 2010 may reflect these evolving treatment approaches.\\


Medical costs associated with breast cancer are often influenced by factors such as cancer sub-type and stage, treatment strategies, and follow-up protocols. Additionally, demographic and regional variations in healthcare access can significantly impact these costs. However, such information is insufficiently captured in Taiwan's claims data, and our analysis provides only an average representation. While the primary aim of this section is to demonstrate how the proposed method can inform multimorbidity management, a potential solution is to link claims data with other sources, such as electronic health records, disease registries, and census data, to enhance the depth and comprehensiveness of analysis.

\section{Conclusion}

We propose a dynamic approach for studying interconnections of multimorbidity-related clinical outcomes. Our model evaluates conditional and time-varying interconnections in disease-specific clinical outcomes, while effectively addressing the issue of zero-inflation. We establish the theoretical properties of the proposed method and validate it through comprehensive simulation studies. Using Taiwan's claims data from 2010 to 2013, we construct a dynamic HDN system on medical costs, comprising a series of temporally correlated networks. Analysis of network structures reveals insights that extend beyond existing literature on human diseases. To illustrate how the constructed network can enhance multimorbidity management, we examine the network structures for breast cancer and discuss the implications. The results of our analysis also have potential applications in other health settings, including treatment strategy development, medical resources allocation, insurance product designing, and health policy formulation.\\

The proposed method can be further developed in several ways. We have focused on medical costs, which is one of the most important considerations in accessing disease burdens. To describe disease interconnections on clinical outcomes measures more comprehensively, more outcomes regarding healthcare quality (e.g., cure, death, recurrence) and efficiency (e.g., the number of visits) could be considered. It is noted that the proposed regression-based model is readily flexible to be expanded incorporating more outcomes upon availability of more data. Moreover, data analysis does not incorporate covariates and reflects population average. The proposed model can be further expanded to incorporate covariates and to examine subgroup (by gender, age, location, etc.) differences. In addition, the proposed network is unweighted and undirected. It may be of interests to further develop the network model to incorporate weighted edges for quantifying magnitude of associations and directed edges for possible causal inference. This may require additional investigations and is defer to future research.

\bmsection*{Acknowledgment}
This work is supported by National Natural Science Foundation of China (72301283) and New Faculty Startup Fund Project at Renmin University of China (23XNKJ07). We extend our heartfelt appreciation to the editor and the reviewers for their insightful comments and valuable suggestions, which have greatly improved the quality of this article.

\clearpage
\bibliography{wileyNJD-AMA.bib}

\bmsection*{Supporting information}

Additional supporting information can be found online in the Supporting Information section at the end of this article.

\clearpage
\begin{table}[htbp]
    \centering
    \caption{Simulation results: average F-score (standard deviation of replications)}
    \label{tab:sim F1}
    \footnotesize
    \resizebox{1\linewidth}{!}{
        \begin{tabular}{cccccccc}
            \hline
            & Sparsity & \multicolumn{3}{c}{5\%} & \multicolumn{3}{c}{15\%} \\ \cmidrule(lr){2-5} \cmidrule(lr){6-8} 
            & Method & Proposed & Alternative1 & Alternative2 & Proposed & Alternative1 & Alternative2\\  \hline
            & $d$ & \multicolumn{6}{c}{Time-varying mechanism 1} \\ \cline{3-8}
            \multirow{3}{*}{Band} & 100 & 0.86 (0.03) & 0.67 (0.05) & 0.08 (0.01) & 0.71 (0.04) & 0.61 (0.05) & 0.11 (0.01) \\
            & 125 & 0.70 (0.05) & 0.60 (0.07) & 0.07 (0.01) & 0.73 (0.04) & 0.69 (0.07) & 0.12 (0.02) \\
            & 150 & 0.72 (0.03) & 0.69 (0.04) & 0.06 (0.02) & 0.80 (0.02) & 0.74 (0.04) & 0.12 (0.01) \\
            \multirow{3}{*}{SBM} & 100 & 0.72 (0.03) & 0.61 (0.04) & 0.17 (0.02) & 0.75 (0.02) & 0.70 (0.02) & 0.13 (0.08) \\
            & 125 & 0.69 (0.03) & 0.61 (0.05) & 0.08 (0.02) & 0.76 (0.03) & 0.66 (0.04) & 0.17 (0.04) \\
            & 150 & 0.72 (0.03) & 0.58 (0.06) & 0.07 (0.01) & 0.76 (0.03) & 0.69 (0.04) & 0.12 (0.02) \\
            \multirow{3}{*}{WS} & 100 & 0.70 (0.04) & 0.63 (0.03) & 0.10 (0.02) & 0.79 (0.03) & 0.73 (0.02) & 0.20 (0.03) \\
            & 125 & 0.72 (0.03) & 0.64 (0.04) & 0.08 (0.02) & 0.78 (0.02) & 0.71 (0.02) & 0.14 (0.01) \\
            & 150 & 0.70 (0.03) & 0.60 (0.06) & 0.06 (0.01) & 0.75 (0.03) & 0.67 (0.06) & 0.12 (0.02) \\
            \cline{3-8}
            & $d$ & \multicolumn{6}{c}{Time-varying mechanism 2} \\ \cline{3-8}
            \multirow{3}{*}{Band} & 100 & 0.73 (0.10) & 0.61 (0.18) & 0.14 (0.06) & 0.74 (0.07) & 0.61 (0.11) & 0.18 (0.05) \\
            & 125 & 0.75 (0.09) & 0.66 (0.17) & 0.16 (0.09) & 0.79 (0.05) & 0.60 (0.15) & 0.19 (0.06) \\
            & 150 & 0.81 (0.06) & 0.74 (0.07) & 0.13 (0.08) & 0.84 (0.02) & 0.76 (0.02) & 0.16 (0.03) \\
            \multirow{3}{*}{SBM} & 100 & 0.72 (0.12) & 0.71 (0.08) & 0.18 (0.10) & 0.78 (0.05) & 0.74 (0.03) & 0.24 (0.05) \\
            & 125 & 0.78 (0.06) & 0.72 (0.08) & 0.17 (0.08) & 0.81 (0.03) & 0.74 (0.03) & 0.18 (0.04) \\
            & 150 & 0.78 (0.06) & 0.70 (0.07) & 0.13 (0.06) & 0.80 (0.03) & 0.73 (0.04) & 0.15 (0.03) \\
            \multirow{3}{*}{WS} & 100 & 0.78 (0.08) & 0.73 (0.08) & 0.17 (0.08) & 0.80 (0.04) & 0.76 (0.03) & 0.24 (0.04) \\
            & 125 & 0.79 (0.05) & 0.72 (0.08) & 0.14 (0.06) & 0.81 (0.03) & 0.74 (0.03) & 0.17 (0.03) \\
            & 150 & 0.77 (0.06) & 0.69 (0.08) & 0.12 (0.06) & 0.81 (0.03) & 0.74 (0.04) & 0.16 (0.03) \\
            \hline
        \end{tabular}
    }
\end{table}

\clearpage
\begin{table}[htbp]
    \centering
    \caption{Simulation results: $\Delta$ and $\Delta_1, \Delta_2$ (standard deviation of replications)}
    \label{tab:sim delta}
    \footnotesize
    \resizebox{1\linewidth}{!}{
        \begin{tabular}{cccccccccc}
            \hline
            & Sparsity & \multicolumn{4}{c}{5\%} & \multicolumn{4}{c}{15\%} \\ 
            \cmidrule(lr){2-6} \cmidrule(lr){7-10} 
            & Method & \multicolumn{2}{c}{Proposed} & \multicolumn{2}{c}{Alternative1}  & \multicolumn{2}{c}{Proposed} & \multicolumn{2}{c}{Alternative1} \\  \hline
            & $d$ & \multicolumn{8}{c}{Time-varying mechanism 1, $\Delta$} \\
            \cline{3-10}
            \multirow{3}{*}{Band} & 100 & \multicolumn{2}{c}{71.57 (9.99)} & \multicolumn{2}{c}{239.76 (17.95)} & \multicolumn{2}{c}{92.88 (16.23)} & \multicolumn{2}{c}{245.66 (16.33)}   \\
            & 125 & \multicolumn{2}{c}{126.73 (20.19)} & \multicolumn{2}{c}{387.06 (17.57)} & \multicolumn{2}{c}{140.85 (32.94)} & \multicolumn{2}{c}{401.26 (40.42)}  \\
            & 150 & \multicolumn{2}{c}{199.67 (43.84)} & \multicolumn{2}{c}{625.74 (47.73)} & \multicolumn{2}{c}{369.83 (73.19)} & \multicolumn{2}{c}{769.58 (109.25)}  \\
            \multirow{3}{*}{SBM} & 100 & \multicolumn{2}{c}{73.00 (10.51)} & \multicolumn{2}{c}{391.65 (21.03)} & \multicolumn{2}{c}{104.12 (19.28)}  & \multicolumn{2}{c}{453.85 (31.06)}   \\
            & 125 & \multicolumn{2}{c}{120.21 (21.47)} & \multicolumn{2}{c}{632.06 (16.84)} & \multicolumn{2}{c}{175.69 (29.67)} & \multicolumn{2}{c}{739.33 (55.90)} \\
            & 150 & \multicolumn{2}{c}{202.19 (39.66)} & \multicolumn{2}{c}{564.02 (45.70)} & \multicolumn{2}{c}{302.19 (73.22)} & \multicolumn{2}{c}{699.39 (86.50)}  \\
            \multirow{3}{*}{WS} & 100 & \multicolumn{2}{c}{73.18 (12.46)} & \multicolumn{2}{c}{224.49 (19.18)} & \multicolumn{2}{c}{123.37 (24.58)} & \multicolumn{2}{c}{299.28 (28.16)}  \\
            & 125 & \multicolumn{2}{c}{123.38 (16.87)} & \multicolumn{2}{c}{393.01 (31.21)} & \multicolumn{2}{c}{217.95 (21.82)} & \multicolumn{2}{c}{455.85 (30.71)}\\
            & 150 & \multicolumn{2}{c}{177.36 (31.63)} & \multicolumn{2}{c}{558.80 (45.68)} & \multicolumn{2}{c}{319.89 (86.51)} & \multicolumn{2}{c}{674.06 (111.70)}  \\
            \cline{3-10}
            & $d$ & \multicolumn{8}{c}{Time-varying mechanism 2} \\
            \cline{3-10}
             &  & $\Delta_1$ & $\Delta_2$ & $\Delta_1$ & $\Delta_2$ & $\Delta_1$ & $\Delta_2$ & $\Delta_1$ & $\Delta_2$ \\
            \cmidrule(lr){3-4} \cmidrule(lr){5-6} \cmidrule(lr){7-8}  \cmidrule(lr){9-10}
            \multirow{3}{*}{Band} & 100 & 191.50 (21.07) & 88.90 (7.63) & 368.40 (20.56) & 266.14 (6.14) & 196.62 (33.02) & 88.03 (11.74) & 542.41 (31.39) & 408.58 (13.67)  \\
            & 125 & 305.60 (32.23) & 137.09 (11.51) & 350.59 (40.52) & 193.31 (11.09) & 325.83 (38.06) & 150.14 (10.25) & 918.12 (44.59) & 666.81 (10.87) \\
            & 150 & 477.98 (19.26) & 209.64 (13.51) & 674.66 (51.92) & 395.54 (12.76) & 713.43 (43.11) & 274.46 (13.81) & 1601.79 (41.23) & 1023.63 (15.89)\\
            \multirow{3}{*}{SBM} & 100 & 183.91 (27.62) & 83.72 (13.40) & 346.74 (31.51) & 247.62 (10.28) & 200.16 (28.14) & 104.08 (8.41) & 582.18 (26.93) & 453.95 (6.93) \\
            & 125 & 287.70 (27.93) & 145.90 (9.77) & 615.42 (35.22) & 418.27 (24.15) & 316.74 (39.49) & 170.59 (7.92) & 865.98 (145.34) & 669.69 (102.84)\\
            & 150 & 421.52 (52.90) & 206.64 (13.82) & 861.83 (39.40) & 618.71 (18.39) & 442.34 (38.73) & 241.81 (15.39) & 1041.78 (197.18) & 766.57 (164.20)\\
            \multirow{3}{*}{WS} & 100 & 188.08 (28.40) & 91.93 (9.56) & 366.12 (31.71) & 250.44 (10.93) & 209.72 (30.91) & 115.44 (11.55) & 408.14 (19.13) & 292.93 (10.09) \\
            & 125 & 308.17 (33.67) & 149.79 (7.26) & 601.98 (42.42) & 408.61 (15.43) & 334.41 (42.42) & 190.70 (15.21) & 653.30 (53.18) & 461.97 (23.04)\\
            & 150 & 406.04 (46.06) & 202.72 (10.97) & 847.08 (66.19) & 589.29 (36.36) & 466.97 (37.20) & 250.55 (13.09) & 956.09 (36.38) & 682.33 (17.47)\\
            \hline
        \end{tabular}
    }
\end{table}

\clearpage
\begin{table}[htbp]
    \centering
    \caption{The number of edges and modules in the networks estimated by different methods}
    \label{tab:edges and modules}
    \footnotesize
        \begin{tabular}{ccccccc}
            \hline
             \multirow{2}{*}{Year} & \multicolumn{2}{c}{Proposed} & \multicolumn{2}{c}{Alternative1} & \multicolumn{2}{c}{Alternative2} \\ 
            \cmidrule(lr){2-3} \cmidrule(lr){4-5} \cmidrule(lr){6-7} 
            & Edges & Modules& Edges & Modules& Edges & Modules \\  \hline
            2000 & 990 & 11 & 1124& 11 &   - & -  \\
            2001 & 911 & 12 & 1190& 12 &  210 &  10 \\
            2002 & 876 &  11 & 1101& 13 &  216  & 10  \\
            2003 & 892 &  11 & 1303& 11 &   210 &  11 \\
            2004 & 894 &  10 & 1374& 10 &  242 &  9 \\
            2005 & 901 &  11 & 1301& 10 &  242 &  11 \\
            2006 & 915 &  13& 1326& 9 &  218 &  11 \\
            2007 & 937 &  13& 1290& 10 &  224  &  10 \\
            2008 & 933 &  12& 1266& 9 &  268 &  8 \\
            2009 & 859 &  10& 1419& 8 &  298  &  10 \\
            2010 & 884 &  9& 1549& 7 &  306 &  9 \\
            2011 & 924 &  12& 1453& 7 &  294 &  11 \\
            2012 & 960 &  12& 1366& 10 & 267  &  12 \\
            2013 & 1093 &  13& 1556& 9 &  335  &  14 \\
            \hline
        \end{tabular}
\end{table}

\clearpage
\begin{table}[htbp]
    \caption{Diseases that are connected or clustered with breast cancer}
    \centering
    \vspace{.3cm}
    \label{tab: breast cancer}
    \resizebox{\textwidth}{!}{%
    \begin{tabular}{llll}
        \toprule
        \multicolumn{4}{l}{\textbf{Diseases that are connected to breast cancer}} \\
        \hline
        CCS Label & Frequency & CCS Label & Frequency \\
        \hline
Other upper respiratory infections &        14 &                                       Disorders of teeth and jaw &        14 \\
  Nonmalignant breast conditions &        14 &                                           Essential hypertension &         7 \\
Diabetes mellitus without complication &         7 &                            Other and unspecified benign neoplasm &         7 \\
       Benign neoplasm of uterus &         6 &                                   Other female genital disorders &         6 \\
     Deficiency and other anemia &         6 &                                            Heart valve disorders &         6 \\
                        Cataract &         5 &   \\
        \hline
        \multicolumn{4}{l}{\textbf{Diseases that are clustered with breast cancer}} \\
        \hline
        CCS Label & Frequency & CCS Label & Frequency \\
        \hline
        Hodgkin`s disease &        14 & Cancer of testis &        14 \\
        Cancer of thyroid &        14 & Cancer of stomach &        13 \\
        Cancer of uterus &        12 & Screening and history of mental health and substance abuse codes &        12 \\
        Disorders of teeth and jaw &        12 & Cancer of other urinary organs &        12 \\
        Other non-epithelial cancer of skin &        12 & Cancer of pancreas &        12 \\
        Leukemias &        12 & Other upper respiratory infections &        11 \\
        Cancer of head and neck &        11 & Cancer of other female genital organs &        11 \\
        Cancer of brain and nervous system &        11 & Cancer of ovary &        11 \\
        Cancer of kidney and renal pelvis &        11 & Cancer of other GI organs; peritoneum &        11 \\
        Cancer of cervix &        11 & Cancer of esophagus &        10 \\
        Melanomas of skin &        10 & Cancer of prostate &        10 \\
        Cancer of rectum and anus &        10 & Cancer of bladder &         9 \\
        Cancer of colon &         9 & Cancer of bone and connective tissue &         8 \\
        Non-Hodgkin`s lymphoma &         8 & Cancer of bronchus; lung &         8 \\
        Diseases of mouth; excluding dental &         7 & Hemorrhoids &         6 \\
        Systemic lupus erythematosus and connective tissue disorders &         6 & Other ear and sense organ disorders &         6 \\
        Neoplasms of unspecified nature or uncertain behavior &         6 & Septicemia (except in labor) &         6 \\
        Paralysis &         5 & Cancer of liver and intrahepatic bile duct &         5 \\
        \bottomrule
    \end{tabular}
    }
\end{table}

\end{document}